\documentclass[%
 reprint,
 amsmath,amssymb,
 aps,
 pra,
]{revtex4-2}

\usepackage{graphicx}% Include figure files
\usepackage{dcolumn}% Align table columns on decimal point
\usepackage{bm}% bold math
\usepackage{braket}
\usepackage{mathtools}
\usepackage{float}
\usepackage{hyperref}
\usepackage{flushend}
\usepackage{xcolor}
\usepackage{afterpage}
\usepackage{placeins}
\begin{document}

\preprint{APS/123-QED}

%\title{A link between Fourier transform spectra and Wigner phase-space representations by continous parametric LE transform}
\title{Surprising applications of Newton's hyperbolism transform of curves in Fourier-transform spectroscopy}
%Conceptual advances in phase-sensitive Fourier transform spectroscopy based on the Newtonian transformation of curves
% Force line breaks with \\

\author{Dennis Huber}
 \altaffiliation{dennis.huber@tum.de}%Lines break automatically or can be forced with \\
\author{Steffen J. Glaser}%
 \email{glaser@tum.de}
% TODO: Update affiliation Layout
\affiliation{Technical University of Munich, School of Natural Sciences, Lichtenbergstrasse 4, 85748 Garching bei M\"unchen, Germany}
\affiliation{Munich Center for Quantum Science and Technology (MCQST), 80799 M\"unchen, Germany}

\date{\today}% It is always \today, today,

\begin{abstract}
\noindent
The Fourier transform (FT) represents a key tool in modern spectroscopy which drastically reduces measurement times and helps to improve the signal-to-noise ratio in spectra. Fourier transforming exponentially decaying time domain signals gives Lorentzian line shapes which can be manipulated by apodization methods.
The underlying transitions of spectral lines can be visualized by a Bloch vector or equivalent phase-space representations.
%On the other hand, representations such as the Bloch vector or phase-space representations are popular tools for visualizing states and processes of quantum systems. 
Here, we study and generalize a surprisingly elegant geometric transform, the hyperbolism of curves originally found by Isaac Newton, which allows to transform ellipses into Lorentzian lines, and vice versa. With this, we show that the Bloch picture and especially corresponding phase-space representations are directly geometrically related to the Lorentzian line shape. We also introduce a novel continuous parametrization of Newton's transform which results in further interesting line shapes. In particular, we find that truncated parabolic lines with finite support can be obtained by the half transform and introduce a new apodization approach to replicate this line shape in experimental spectra. We discuss concrete applications in nuclear magnetic resonance spectroscopy.

%\begin{description}
%\end{description}
\end{abstract}

\keywords{Spectroscopy, Signal processing, Nuclear magnetic resonance, Phase space methods}

%Use showkeys class option if keyword
\maketitle

%\tableofcontents

\section{Introduction}\label{sec:intro}
\setlength{\parskip}{0cm}
\noindent
In Fourier-transform (FT) spectroscopy, Lorentzian lines are obtained from exponentially decaying time domain signals. This signal decay is invoked by relaxation processes \citep{Bloch, Keeler} and leads to line broadening and to the characteristic Lorentzian line shape in the frequency domain, i.e., the spectrum.

Historically, Lorentzians have been known and studied for many centuries albeit under different names \citep{NamesAgnesi}. Seminal works on the curve can be traced back to Pierre de Fermat \citep{Fermat}, Isaac Newton \citep{Newton}, Guido Grandi \citep{Grandi}, and Maria Gaetana Agnesi \citep{Agnesi}.  A mistranslation of Agnesi's publication later lead to the curve being famously known as the \textit{Witch of Agnesi} \citep{NamesAgnesi}. Almost a century later, the curve was studied as a probability density function by Poisson \citep{Poisson}. The function was associated with Augustin Cauchy \citep{Stigler} almost three decades later and became known as the \textit{Cauchy distribution} in mathematics whereas the term Lorentz distribution or \textit{Lorentzian function}, named after Hendrik Lorentz, became popular among physicists.

Lorentzian lines have more pronounced tails in comparison to other important line shapes such as Gaussians of the same width (full width at half maximum, FWHM). Previously, this served as a motivation for the development of signal apodization techniques \citep{Keeler, Derome, Hoch, Bodenhausen} that allow to manipulate the spectral line shapes post-measurement. Technically, this is achieved by multiplying the time domain signal with window functions that alter the envelope curve shape in a way such that the desired spectral line shapes are obtained.\newline
A prominent example of such apodization methods is the \textit{Lorentz-to-Gaussian} transform \citep{Keeler, Derome, Hoch} that allows to obtain narrower Gaussians lines while lowering the signal-to-noise ratio.

\begin{figure*}[t!]
\includegraphics[width=.95\textwidth]{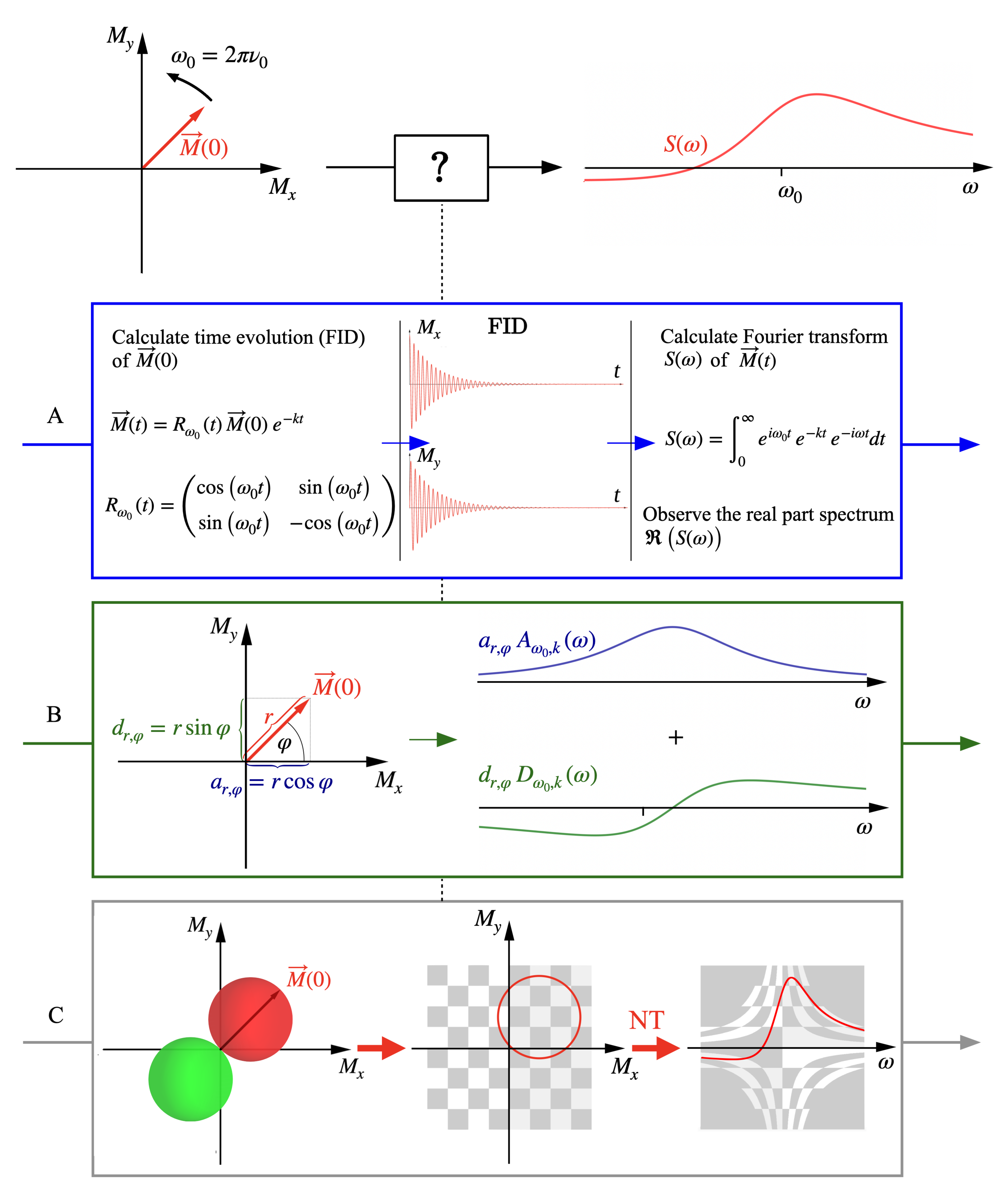}
\caption{\label{fig:IntroFig}In magnetic resonance spectroscopy, being given an initial transverse magnetization vector $\vec{M}(0)$ that evolves with a frequency $\omega_0$ and transverse relaxation time $T_2$, phase-sensitive spectral lines can be obtained (\textbf{A}) by calculating the time evolution $\vec{M}(t)$ of $\vec{M}(0)$ to obtain the free induction decay (FID) which can be Fourier transformed. This gives a complex spectrum of which only the real part is commonly observed. (\textbf{B}) alternatively, it is straightforward to analyze the initial transverse magnetization components of $\vec{M}(0)$ which can be mapped to pure absorption or dispersion mode Lorentzians that, when combined directly, give the spectral line shape. (\textbf{C}) circles corresponding to the intersection of the phase-space representation \citep{DROPS, BEADS} in polar form \citep{LeinerWQST} of $\vec{M}(0)$ can be directly transformed to corresponding spectral lines by the so-called hyperbolism or Newton transform (NT) of curves as will be shown in the following.}
\end{figure*}

As shown in Fig.~\ref{fig:IntroFig}, in nuclear magnetic resonance spectroscopy, given an initial transverse magnetization vector (Bloch vector \citep{Bloch,Feynman}) $\vec{M}(0)$ of an observable single spin transition \citep{Bodenhausen} with frequency $\nu_0$ and the transverse relaxation damping constant $k =1/T_2$ 
\begin{equation}\label{eq:Bloch}
\vec{M}(0) = \begin{pmatrix}
M_x(0)\\
M_y(0)\\
0
\end{pmatrix},
\end{equation}
the line shape of the resulting spectrum, if the free induction decay (FID) of the detectable time-domain signal could be observed during an infinitely long detection period, can be derived in various ways.

One approach (\textbf{A} in Fig.~\ref{fig:IntroFig}) to obtain the line shape is given by first calculating the time evolution of the system which is then Fourier transformed to give a spectrum.

However, the most direct and simple method to derive the line shape involves the linear combination of pure absorption $A_{\omega_0, k}(\omega)$ and dispersion mode $D_{\omega_0, k}(\omega)$ Lorentzian lines weighted by the corresponding components of the Bloch vector $\vec{M}(0)$ (\textbf{B} in Fig.~\ref{fig:IntroFig}). Here, we assume the standard case where the detection setup is chosen such that $M_x(0) = a_{r,\varphi} = r\cos\varphi$ (using the phase $\varphi$ and $r$, the length of $\vec{M}(0)$) is assigned to the absorption mode component and $M_y(0) = d_{r,\varphi} = r\sin\varphi$ corresponds to the dispersion mode Lorentzian component. 

\clearpage

\noindent
The spectral line is then obtained as
\begin{align}
S_{\omega_0, k, \varphi, r}(\omega) = a_{r,\varphi} A_{\omega_0, k}(\omega) &+ d_{r,\varphi} D_{\omega_0, k}(\omega)\nonumber \\
= r\cos\varphi  \frac{k}{(\omega - \omega_0)^2 + k^2} &-  r\sin\varphi \frac{(\omega - \omega_0)}{(\omega - \omega_0)^2 + k^2}.
\end{align}

\noindent
An alternative way to represent a Bloch vector (Eq.~\ref{eq:Bloch}) is given by continuous phase-space representations, e.g., the DROPS \citep{DROPS}, BEADS \citep{BEADS} and PROPS representations \citep{PROPS}. The DROPS representation is particularly relevant for visualizing coupled spin systems in nuclear magnetic resonance as it conserves and directly represents properties such as symmetries with respect to permutation or rotation \citep{DROPS} and it was even shown to be experimentally measurable by Wigner state or process tomography \citep{LeinerWQST, LeinerWQPT, Devra}. Indeed, a simple phase-space description of a single-transition vector corresponds to a function which consists of two equally sized spheres of opposite sign (a so-called real spherical harmonic of rank $j=1$ \cite{Whittaker}). The transverse Bloch vector $\vec{M}(0)$ can then be visualized in a simplified fashion as the intersection circles of these two spheres with the transverse plane (see \textbf{C} in Fig.~\ref{fig:IntroFig}). Being given this representation, we wondered whether a direct geometric relation between this representation and phase-sensitive Lorentzian line shapes in corresponding spectra exists.

Indeed, we found a suprisingly elegant solution to this problem and after digging deeper in the mathematical literature we noticed that we had rediscovered a geometric transformation called the \textit{ Newton hyperbolism transform} of curves named after Isaac Newton who first proposed the underlying geometric relations in his work on Lorentzian curves, which was only published posthumously in 1779 \citep{Newton} and independently found by Grandi~\citep{Grandi}.

In the following, we will thus introduce the hyperbolism transform by an illustrative mathematical-geometric model and show how it can be applied to determine spectral line shapes in section~\ref{sec:ch1}. We then proceed to generalize the Newton transform of ellipses by introducing a parametrization in section~\ref{sec:ch3} which allows to continuously transform between ellipses and Lorentzians. The appearance of spectroscopic line shapes from this unconventional perspective provides interesting new insights: Absorption mode truncated parabolic lines can be found, that have finite support (section~\ref{sec:ch4}) and which can be used to obtain a simplified spectral representation. Motivated by these findings, the corresponding window function for a Lorentzian-to-truncated parabola transform is derived in section~\ref{sec:ch5} that allows to realize finite support truncated parabolic line shapes in an experimental context. Experimental tests and theoretical studies are presented which examine the effect of incomplete compensation of exponential decays and the effects of noise on the new Lorentzian-to-truncated parabola transform. In addition, a particularly simple representation of purely dispersive signals which is given by semi-ellipses is introduced.

\section{The Newton Hyperbolism transform}\label{sec:ch1}
\noindent
We first outline the details of how to use the Newton hyperbolism transform of a circle to obtain phase-sensitive Lorentzian lines from a Bloch vector on a purely geometric level whilst initially neglecting physical properties. Afterwards, we will present a step-by-step instruction that reveals how to transform circles into Lorentzian lines in a physical context.

Conceptually, we can imagine the hyperbolism transform as a limiting case of the \textit{rubber sheet geometry} model that is commonly used in topology \citep{Topology}. In this model, transformations are imagined as deformation of a flexible rubber sheet. The Newton hyperbolism transform then amounts to pulling at a square rubber sheet at the horizontal axis (x-axis) with infinite force while holding it at the centers of each side. However, a geometric peculiarity distinguishes this transform from classical continuous rubber sheet geometry: The lower half-plane of the sheet has to be mirrored across the central vertical axis (y-axis).

\begin{figure*}[htb!]
\centering
\includegraphics[width=0.7\textwidth]{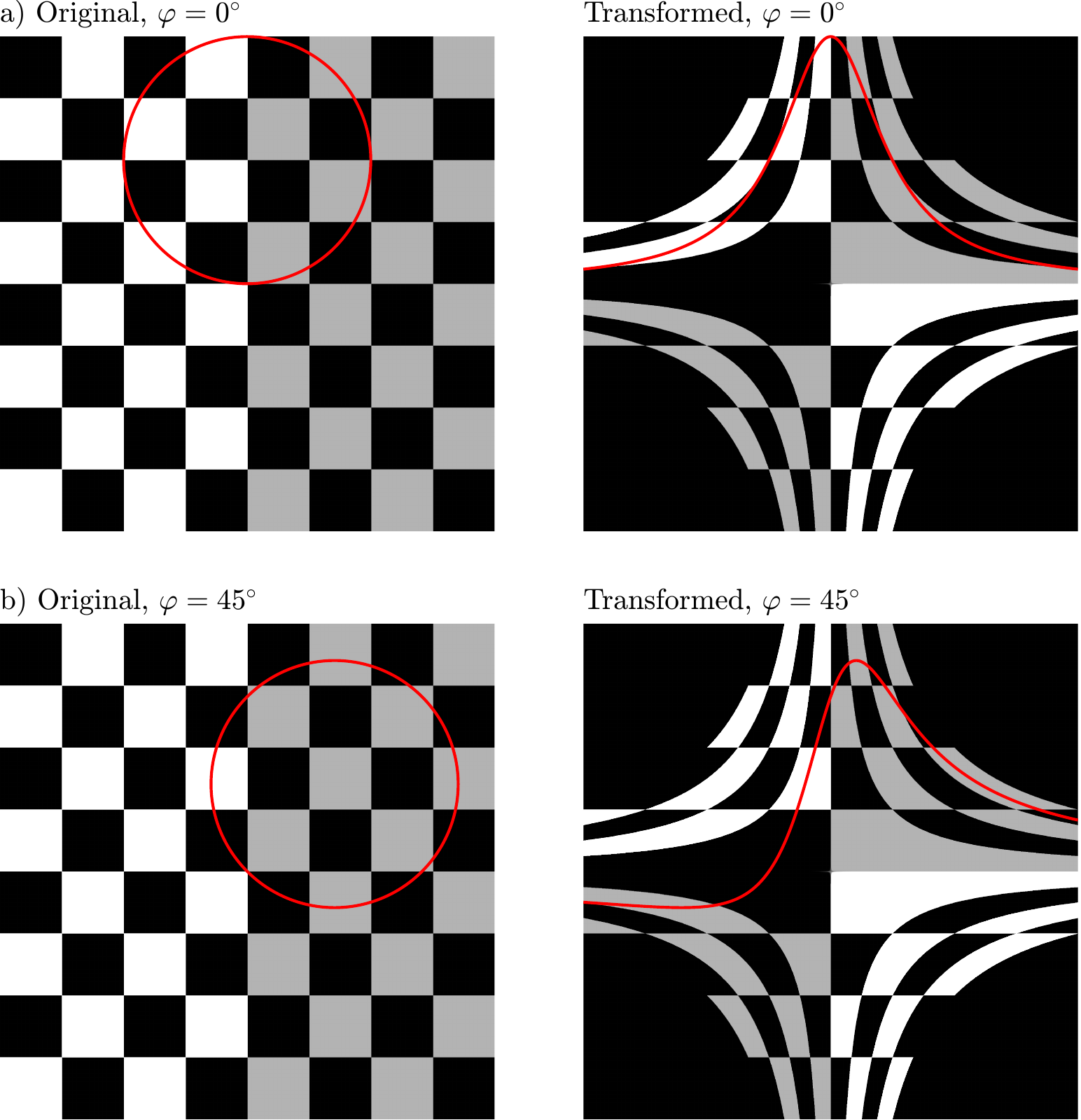}
\caption{\label{RubSheet} Schematic representation of the so-called hyperbolism or Newton transform of curves illustrated by (cut and half-inverted) rubber sheet geometry before and after applying the required deformation. In case (a), a red circle drawn centered in the upper half-plane gives a pure absorption mode Lorentzian ($\varphi = 0^\circ$), whereas the curve obtained in case (b) corresponds to an equal mixture of absorption and dispersion Lorentzian modes  ($\varphi = 45^\circ$). The checkerboard reveals the applied stretchings. The right half-plane in the original picture has grey instead of white squares to illustrate the included lower half-plane mirroring operation.}
\vspace{0cm}
\end{figure*}

As shown in FIG.~\ref{RubSheet}, the induced deformation then causes points to be stretched horizontally in dependence of their position on the sheet. When starting from a circle drawn on the rubber sheet, this yields a Lorentzian line.

The transform can also be written as a single-step mapping $(x,y) \mapsto (\frac{x}{y},y)$. We now present a multi-step procedure to apply the transform in a spectroscopic context and start by observing detectable transverse Cartesian components of examined single-transitions corresponding to single-transition operators $I^{k,l}$ between to energy levels $k$ and $l$ of a system.\citep{Bodenhausen} 

\begin{figure*}[t!]
\includegraphics[width=.84\textwidth]{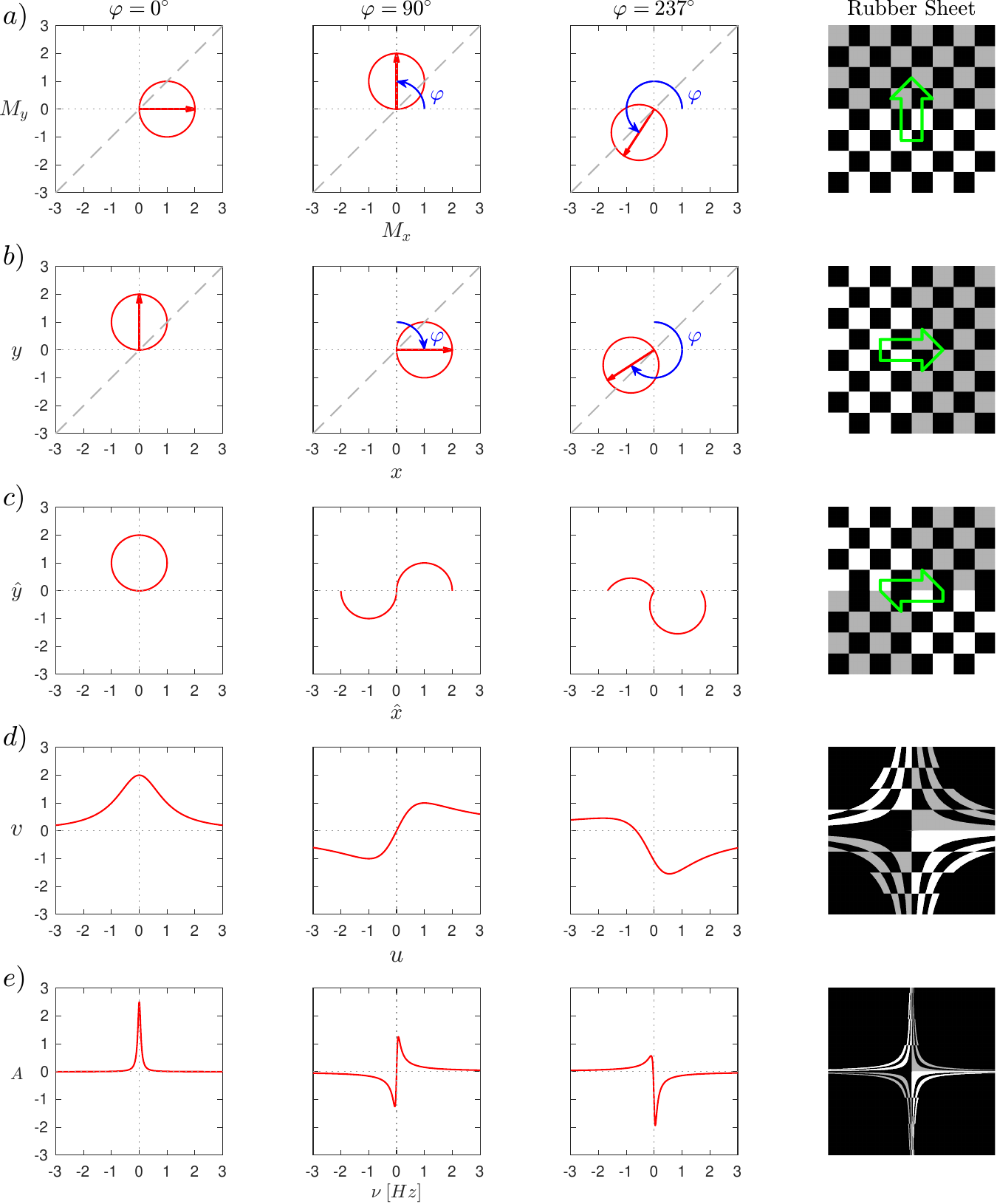}% Here is how to import EPS art
\caption{\label{fig:MainProc}Full Newton hyperbolism transform protocol for three examples of transverse Bloch vectors corresponding to an uncoupled spin or to multiplet components (single-transitions) of coupled spins. The shown Bloch vectors correspond to phase angles $\varphi = 0^\circ$, $90^\circ$ and $237^\circ$. The transform procedure is illustrated using the previously introduced checkerboard image (right column) to clarify the required geometric operations. (a) the Bloch vector and an equivalent spherical function represented by a circle encompassing the vector are drawn into a coordinate system with axes corresponding to the Cartesian components $M_x$ and $M_y$ of the examined single-transition. (b) the coordinates are transformed according to the chosen detector setup by a reflection across the indicated dashed diagonal. Note that for the transformation from step (a) to (b), we exemplarily assume the case where a real-part time-domain signal is detected along the x-axis, and the imaginary part is detected along the y-axis. The lower half-plane is then (c) reflected through the y-axis giving a cartoon like spectrum composed of circular arcs. (d) the latter is then transformed into Lorentzian lines as introduced in eq.~\ref{eq:S4}. (e) finally, physically relevant parameters such as the damping constant and the amplitude (corresponding to the length of the projected Bloch vector), can be introduced by associated scaling in both coordinate dimensions which gives a physically accurate spectral line with correct line shapes as specified in Eq.~\ref{eq:S6}. Here, $r = 0.5$ and $k = 0.4$~Hz. Each step is schematically visualized by a corresponding checkerboard. Green arrows illustrate the applied transformations in steps (a) to (c).}
\end{figure*}

A visual overview of the described procedure using differing single-transition states is given in FIG.~\ref{fig:MainProc}. \newline

\textit{Step 1:} We represent the transverse single-transition components by a corresponding Bloch vector
\begin{eqnarray}
\label{eq:SingleTransition}
I^{k,l} = \begin{pmatrix}
I^{k,l}_x\\
I^{k,l}_y
\end{pmatrix} \longleftrightarrow\vec{M}^{k,l} = \begin{pmatrix}
M^{k,l}_x\\
M^{k,l}_y
\end{pmatrix} 
= 2 \begin{pmatrix}
\braket{I^{k,l}_x}\\
\braket{I^{k,l}_y}
\end{pmatrix},
\end{eqnarray}

\noindent
where $\braket{I^{k,l}_\epsilon} = c^{k,l}_\epsilon = \text{Tr} (\rho\:I^{k,l}_\epsilon)$ is the expectation value of the operator component $I^{k,l}_\epsilon$ ($\epsilon \in \{x,y\}$) for a given density operator $\rho$ that describes the current state of the observed quantum system.

Note that such a vector is always equivalent to a phase-space representation \citep{BEADS} which is a spherical function $f$ composed of rank $j=1$ real spherical harmonics $Y_{1,m}$~\citep{Whittaker} with orders $m \in \{-1,1\}$, i.e.,
\begin{eqnarray}
\label{eq:Wig}
I^{k,l} &&= \sum_{\epsilon \in \{x,y\}} \:c^{k,l}_\epsilon I^{k,l}_\epsilon\nonumber\\ \longleftrightarrow f &&= 2\:\sqrt{\frac{4\pi}{3}}\left(c^{k,l}_{x}Y_{1,1}+ c^{k,l}_{y}Y_{1,-1}\right).
\end{eqnarray}
\vspace{.5cm}

\begin{figure}[h!]
\includegraphics[width=0.46\textwidth]{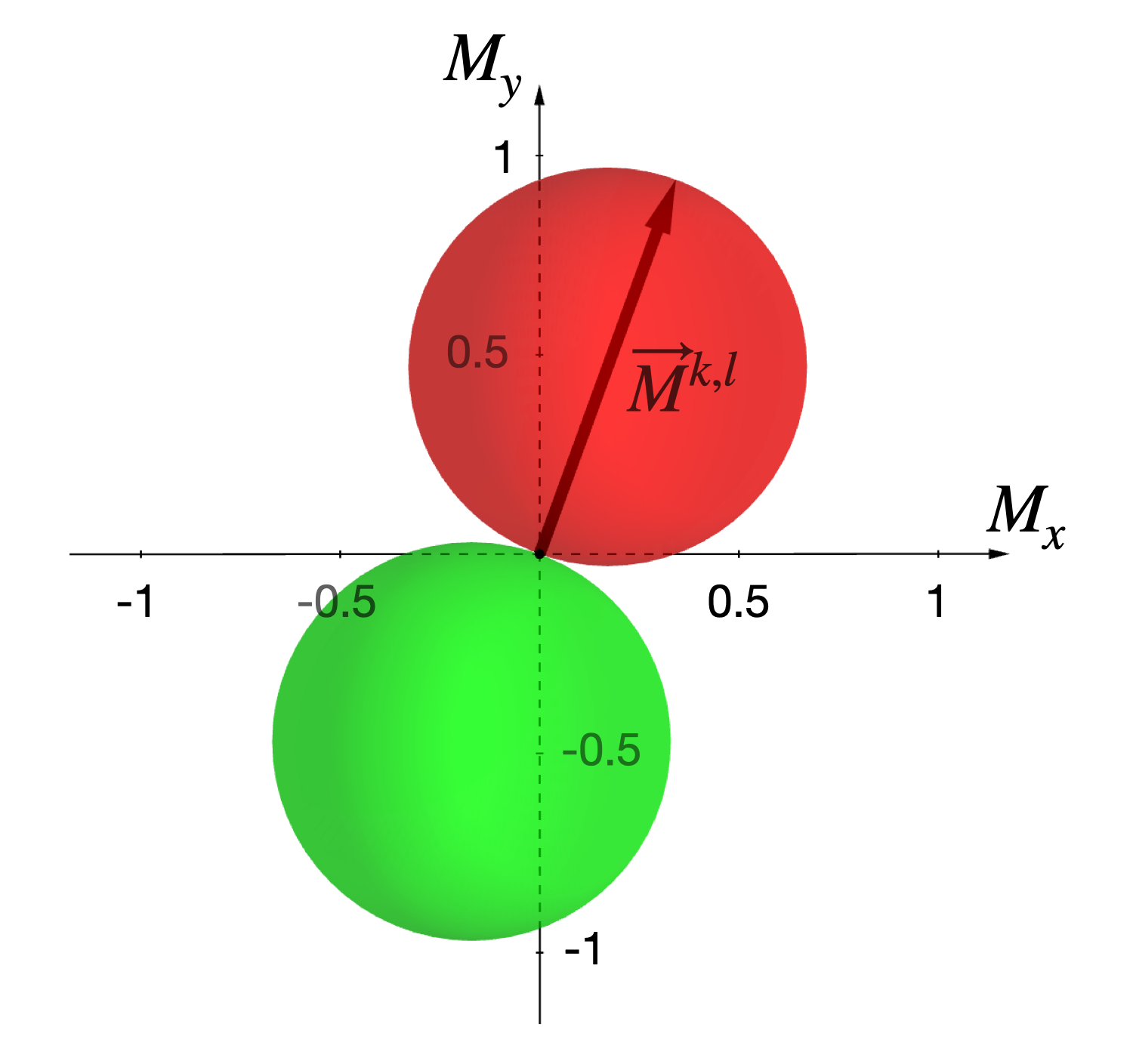}
\caption{\label{STMag} Any single-transition operator $I^{k,l}$ can be represented as a Bloch vector $\vec{M}^{k,l}$ or an equivalent phase-space representation given by a spherical function (red and green shape).}
\vspace{-.6cm}
\end{figure}

\textit{Step 2:} The phase-space representation can be simplified to a circle with the Bloch vector $\vec{M}^{k,l}$ as its diameter that is obtained as the intersection of the original red lobe of the spherical function (see Fig.~\ref{STMag}) and the transverse plane. Note that the negative lobe (green color in Fig.~\ref{STMag}) can be disregarded as the required information is fully encoded in the positive red lobe. All geometric objects can be drawn in a coordinate system with axes $(M_x,M_y)$ corresponding to the transition Bloch vector components. For simplicity, we scale the vector and the encompassing circle such that the circle radius becomes $r = 1$ (corresponding to a vector of length 2):
\begin{eqnarray}
\label{eq:2D}
\vec{M}^{k,l} &&= 2\sqrt{\left(M^{k,l}_x\right)^2 + \left(M^{k,l}_y\right)^2}^{-1}\begin{pmatrix}
M^{k,l}_x\\
M^{k,l}_y
\end{pmatrix}.
\end{eqnarray} 
The scaled projected Bloch vector can be disregarded for the subsequent steps (see Fig.~\ref{fig:MainProc}(a)).

\textit{Step 3:} In order to obtain the correct Lorentzian mode for each transition component, we transform the coordinates $(M_x,M_y)$ in dependence of the quadrature detector setup that is to be replicated.  The transition component that will give the pure absorption mode Lorentzian – usually this corresponds to the real part detector – is assigned to the y-axis and the dispersion mode component to the x-axis of the transformed coordinate system. For instance, if a setup comprises real part detection along the x-axis and imaginary part detection along the y-axis, the vector coordinates chosen in \textit{Step 2} transform into
\begin{eqnarray}
\label{eq:S3}
&&M_x \rightarrow y, \nonumber \\
&&M_y \rightarrow x,
\end{eqnarray}
as shown in Fig.~\ref{fig:MainProc}(b). We will assume this setup for all examples and parametric expressions in the following chapters.
Note that in nuclear magnetic resonance, the chemical shift (ppm) axis is usually plotted from right to left. However, the corresponding frequency axis in units of Hertz for spins with positive gyromagnetic ratio $\gamma > 0$ (e.g., $^1$H or $^{13}$C) increases from left to right \citep{Levitt}. The opposite applies for negative gyromagnetic ratios for which the required coordinate transform becomes $-M_y \rightarrow x$.
\newline

\vspace{-.1cm}\textit{Step 4:} We now perform the first of two steps to achieve the actual Newton transform by reflecting the third and fourth quadrants of the coordinate system across the y-axis. We get
\begin{eqnarray}
\label{eq:S4}
&&x \rightarrow \hat{x} = \text{sgn}(y)\:x, \nonumber \\
&&y \rightarrow \hat{y} = y,
\end{eqnarray}
see Fig.~\ref{fig:MainProc}(c). This will cause points on the circle determined in \textit{Step~2} with negative y-coordinates to be mirrored across the y-axis which ultimately gives a cartoon-like spectrum (see step (c) in FIG.~\ref{fig:MainProc}) in which the line shape is already roughly relatable.
\newline

\vspace{-.1cm}\textit{Step 5:} The defining step of the procedure requires another coordinate transformation (the rubber sheet transformation discussed in Fig.~\ref{RubSheet}).
\begin{eqnarray}
\label{eq:S5}
&&\hat{x} \rightarrow u = \frac{\hat{x}}{|\hat{y}|}, \nonumber \\
&&\hat{y} \rightarrow v = \hat{y}.
\end{eqnarray}
Indeed, this will geometrically transform the previously obtained circular arcs into their hyperbolism, that is, continuous normalized Lorentzian functions with correct phase information corresponding to the line shape of the examined single-transition in an FT spectrum, see Fig.~\ref{fig:MainProc}(d).

\textit{Step 6:} As a last step, one may introduce scalings of the derived Lorentzian lines in both dimensions to adequately represent physical properties such as the damping constant $k$ that is required to describe the exponential decay of a spectroscopically measured time-domain signal, usually expressed by a decay function $f(t)=\exp{(-kt)}$, and the amplitude which is determined by the radius 
\begin{equation}\label{eq:radius}
{r=\frac{\sqrt{(M^{k,l}_x)^2+(M^{k,l}_y)^2}}{2}}
\end{equation}
of the circle encompassing the original unscaled Bloch vector. The physically parameterized Lorentzian mode functions then read
\begin{eqnarray}
\label{eq:AbsLorPar}
A(\nu) = \frac{2rk}{(2\pi\nu)^2+k^2},\\
\label{eq:DispLorPar}
D(\nu) = \frac{2r(2\pi\nu)}{(2\pi\nu)^2+k^2},
\end{eqnarray}
in which case the pure absorption mode Lorentzian $A(\nu)$ adopts a maximum intensity of $A_{max}=\frac{2r}{k}$ and a full width at half maximum $\text{FWHM}=\frac{k}{\pi}$ which implies a coordinate scaling of
\begin{eqnarray}
\label{eq:S6}
&&u \rightarrow \nu = \frac{k}{2\pi}\:u \nonumber, \\
&&v \rightarrow A = \frac{r}{k} v,
\end{eqnarray}
see Fig.~\ref{fig:MainProc}(f). It should be stressed that exact amplitude scaling is not mandatory for most spectroscopic techniques, e.g., nuclear or electron spin resonance spectroscopy, where typically only relative quantities are of interest. Further, the frequency $\nu$ can now be expressed in units of Hertz (Hz).

In Appendix~\ref{AppEll}, we discuss parameterizations of ellipses which entail physical properties and thus directly lead to correctly scaled Lorentzians upon performing the introduced Newton transform omitting \textit{Step 6} of the described procedure.

\section{Parametric continuous transform}\label{sec:ch3}
\noindent
As briefly indicated at the end of section~\ref{sec:ch1}, our transform is generally applicable to ellipses for which circles of course are a special case. We now present a, to our knowledge, new generalized parametrization of the Newton hyperbolism transform using the transform parameter $p \in [0;1]$.
This replaces the transformation in \textit{Step 4} and \textit{5} by the generalized parametric continuous Newton transform of any point $P_p = (x,y)$ in a local coordinate system
\vspace{-.3cm}
\begin{equation}
\label{eq:Abs2CircPar}
(x,y) \mapsto \left(\zeta\eta_E\frac{x}{|y|^p},y\right),
\end{equation}
with $\zeta = \text{sgn}(y)$ as introduced in Step~4 of section~\ref{sec:ch1}. 

\clearpage

\begin{figure*}[htb!]
\includegraphics[width=.9\textwidth]{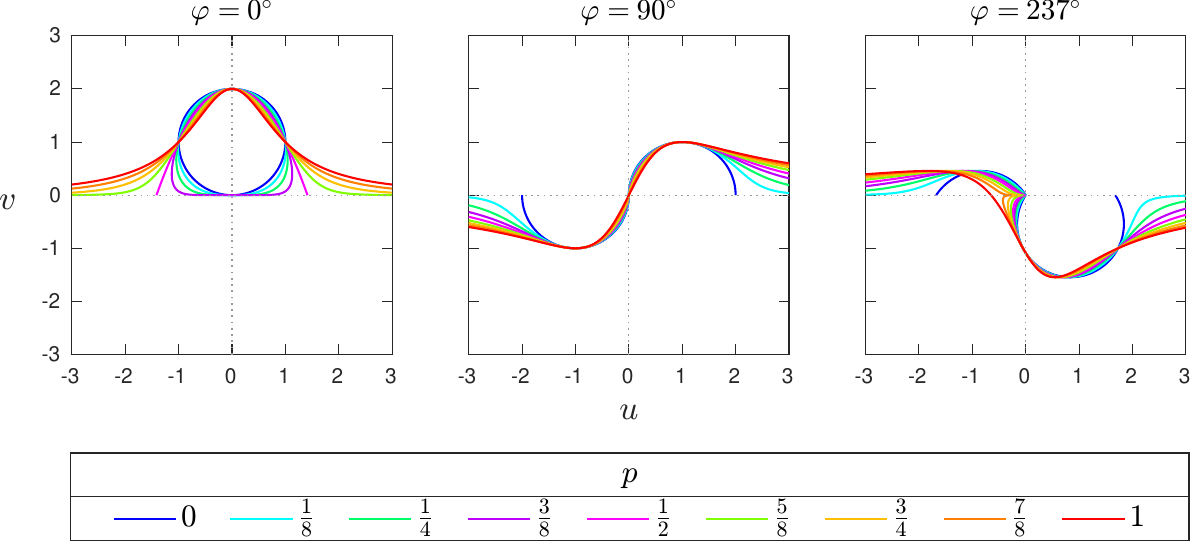}% Here is how to import EPS art
\caption{\label{fig:MainCont}Continuous parametric transform of circles into associated Lorentzians with a transform parameter $p \in [0;1]$ incremented in steps of $\Delta p = \frac{1}{8}$. Read from left to right, the phase angles $\varphi$ defining the position of the circle correspond to a pure absorption mode ($\varphi = 0^\circ$), a pure dispersion mode ($\varphi = 90^\circ$) and an arbitrary case ($\varphi = 237^\circ$). The presented figures show the resulting line shapes for the denoted values of $p$. Grey dotted lines mark the coordinate axes.}
\end{figure*}
\noindent
$\eta_E$ denotes an additional scaling factor required for continuously transforming points of an arbitrarily parametrized ellipse into the same Lorentzian. Several parametrizations of ellipses are described in Appendix~\ref{AppEll} which conserve particular physical properties. In the following, we will use parametrization~A from Appendix~\ref{AppEll}, i.e., ellipses which conserve the full width at half maximum (FWHM) and amplitude of the corresponding absorption mode Lorentzian for our visualizations. 
Referring back to \textit{Step~5} of section~\ref{sec:ch1}, the new continuous Newton transform amounts to a coordinate transform
\begin{eqnarray}
\label{eq:S7mod}
&&\hat{x} \rightarrow u = \frac{\hat{x}}{|\hat{y}|^p}, \nonumber \\
&&\hat{y} \rightarrow v = \hat{y}
\end{eqnarray}
in \textit{Step~5} of the transform protocol, replacing the full transform introduced in Eq.~\ref{eq:S5}.\newline
For pure absorption mode cases ($\varphi \mod \pi = 0$), the bounding values of the transform parameter $p$ yield an ellipse (${p=0}$) or the corresponding Lorentzian ($p=1$). On the other hand, the pure dispersion mode case ($\varphi \mod \pi = \frac{\pi}{2}$) gives split but equally sized semi-ellipses (${p=0}$) and the corresponding Lorentzian ($p=1$).

Exemplary continuous transforms of ellipses into their associated Lorentzians in dependence of the parameter $p$ are shown in Fig.~\ref{fig:MainCont}. The obtained curves differ significantly with respect to their mathematical properties depending on the phase angle~$\varphi$. For pure absorption mode cases, values $0 \leq p < 0.5$ give rotund shapes, for $p = 0.5$ the obtained line is parabolic, and $0.5 < p \leq 1$ gives curves that can be described by mathematical functions.

On the other hand, in pure dispersion mode cases, we obtain semi-ellipses for $p = 0$ and curves that can be described by mathematical functions otherwise.

Any other phase $\varphi$ gives line shapes that cannot be described by mathematical functions for transform parameters $p < 1$.\newline
We present and discuss continuous transforms of parameterized ellipses,  which allow to model physical and mathematical properties, in Appendix~\ref{AppEll}. Note that in general, it is possible for the transform parameter $p$ to also take on values $p<0$ or $p>1$. The transform behavior for such cases is briefly discussed in Appendix~\ref{AppE}.

\section{Spectral lines with finite support}\label{sec:ch4}
\noindent
Being given the generalized continuous Newton transform (see section~\ref{sec:ch3}), we now show that geometrically related spectral line shapes exist which have narrow linewidths and decrease more rapidly away from their extrema than Gaussian lines for x-values outside of the FWHM range. 

\subsection{Absorption mode Lorentzians}\label{sec:AbsLorSup} 
\noindent
As demonstrated in section~\ref{sec:ch3}, the parametric continuous Newton transform allows us to obtain different curves for arbitrary transform parameter values $0<p<1$. In the absorption mode case, the case $p=0.5$ gives truncated parabolic line shapes with finite support.

\begin{figure}[H]
\centering
\hspace{-.3cm}
\includegraphics[width=0.4\textwidth]{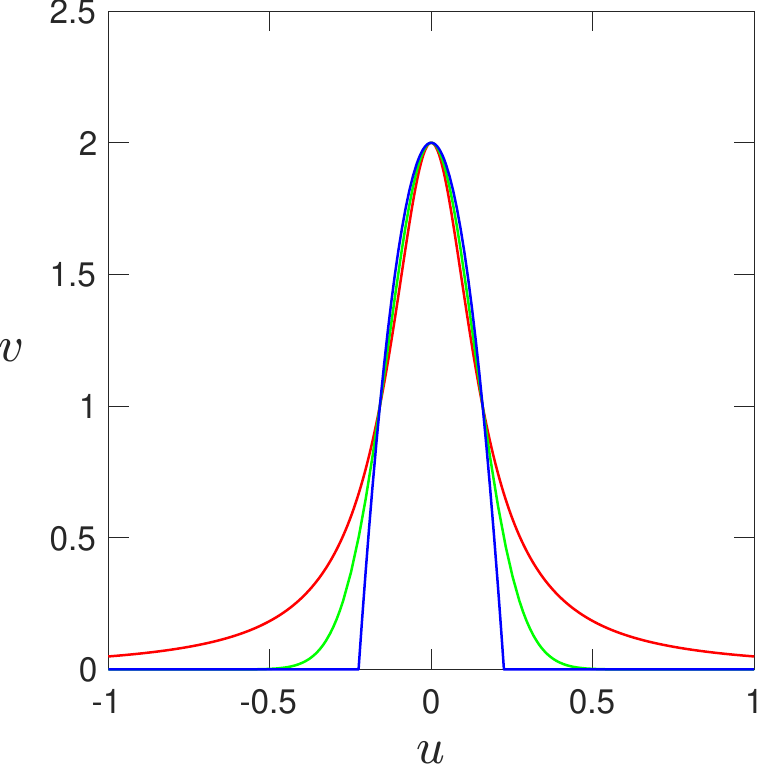}
\caption{\label{fig:ParabolaFWHM}Lorentzian (red), Gaussian (green) and truncated Parabola (blue) with identical maxima and FWHM. All curves were calculated for parametric values $k = 1$~Hz and $r = 1$. Unlike Lorentzians and Gaussians, the truncated parabola has finite support and possesses a comparatively narrow base width.}
\end{figure}

By finding suitable scaling factors $\eta_E$ (see Eq.~\ref{eq:Abs2CircPar}) we can find parametric expressions of parabolic functions which conserve important properties of the original Lorentzian. For instance, the full width at half maximum (FWHM) is conserved by parabolic curves of the form
\begin{eqnarray}
\label{eq:FWHMParab}
&&P_{\text{parab,FWHM}} = \nonumber\\
&&\Bigg(\frac{\frac{\sqrt{rk}}{2\pi}\cos(\theta)}{\sqrt{\frac{r}{k}\left[1+\sin(\theta)\right]}},\frac{r}{k}\left[1+\sin(\theta)\right]\Bigg),
\end{eqnarray}
which can be rewritten as a parametric function
%\begin{equation}
%\label{eq:FWHMParab2}
%v_{\text{FWHM}}(u)=\Bigg(\frac{-4\pi^2 ru^2}{k^3} +\frac{2r}{k}\Bigg)\frac{\text{sgn}\big(\frac{-4\pi^2 ru^2}{k^3} +\frac{2r}{k}\big)+1}{2}
%\end{equation}
\begin{equation}
\label{eq:FWHMParab2}
v_{\text{FWHM}}(u)=\bigg(\frac{-4\pi^2 ru^2}{k^3} +\frac{2r}{k}\bigg)\ H\bigg(\frac{-4\pi^2 ru^2}{k^3} +\frac{2r}{k}\bigg)
\end{equation}
using the radius $r$, the damping constant $k$ (see \ref{sec:ch1} for further details), and the Heaviside function $H$. We provide a comparison of a Lorentzian, a Gaussian of the form $G(u) = \frac{2r}{k}\exp{\big(-\frac{u^2}{2\sigma^2}\big)}$ with $\sigma = \frac{k}{2\pi \sqrt{2\log2}}$, and a truncated parabolic line that possess identical maxima and FWHM in Fig.~\ref{fig:ParabolaFWHM}. It is clear that the truncated parabola has finite support and decreases significantly faster than the Gaussian for x-values outside of the FWHM range and is thus considerably less likely to produce overlaps when used to visualize a spectrum.

\begin{figure}[H]
\centering
\vspace{.1cm}
\includegraphics[width=0.36\textwidth]{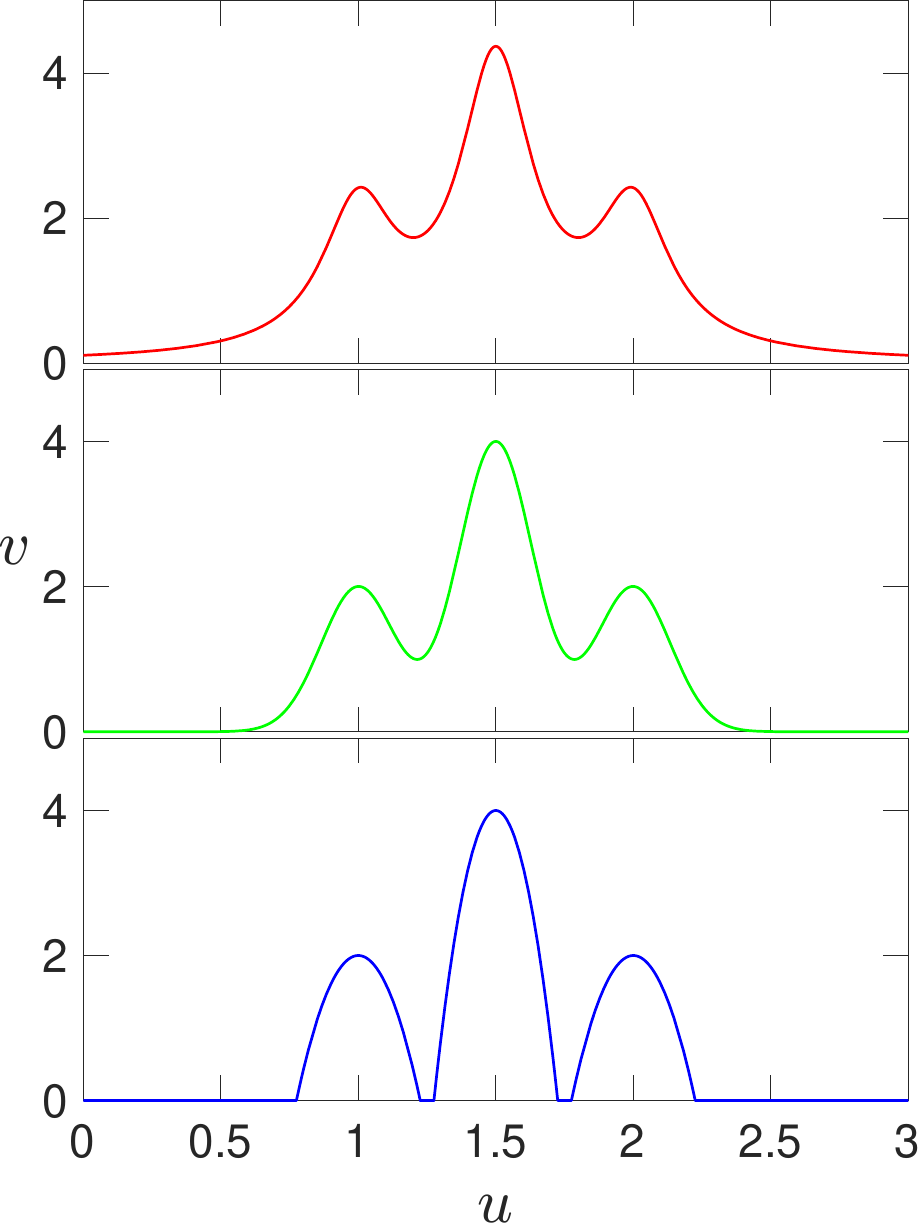}
\caption{\label{fig:FWHMParabolaSpec}Simulated triplet with Lorentzian (red, top), Gaussian (green, center) and truncated parabolic (blue, bottom) line shapes. All variants share the same FWHM and were calculated using a damping constant $k = 1$. In this case, where the line separation is larger than $2\sqrt{2}\:$FWHM, the parabolic lines do not overlap at all whereas the remaining functions do significantly.}
\end{figure}

This is further elucidated in FIG.~\ref{fig:FWHMParabolaSpec} where a simulated triplet is shown for the discussed types of functions. Clearly, the Lorentzian and Gaussian lines overlap significantly. Conversely, the parabolic lines are fully separated.

It should be stressed, that the introduced truncated parabolas do not conserve the original Lorentzian integral area. As discussed before, this is however not required, as spectroscopic applications usually focus on comparing relative quantities. 

\begin{figure}[H]
\hspace{-.3cm}
\begin{minipage}[c]{.5\textwidth}
\hspace{.48cm}
\includegraphics[width=.92 \textwidth]{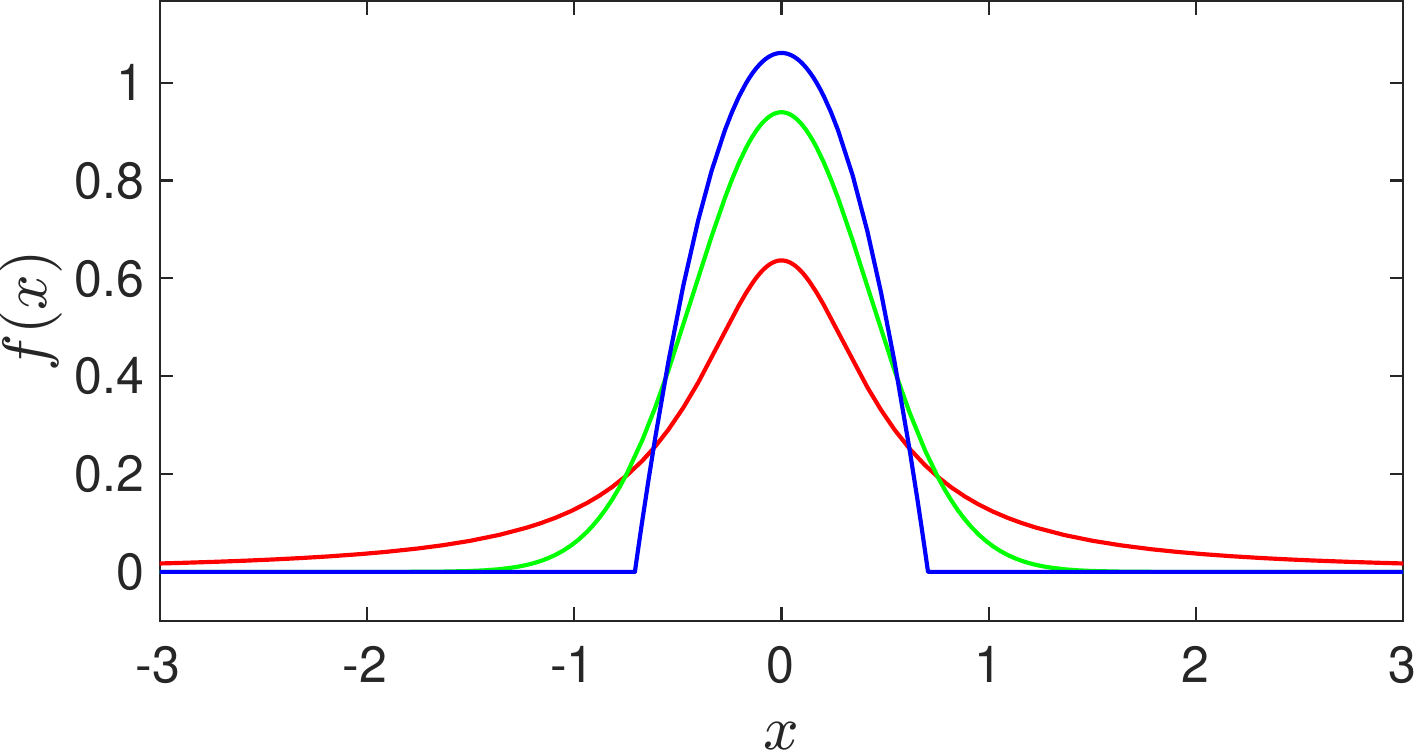}\\
\includegraphics[width=\textwidth]{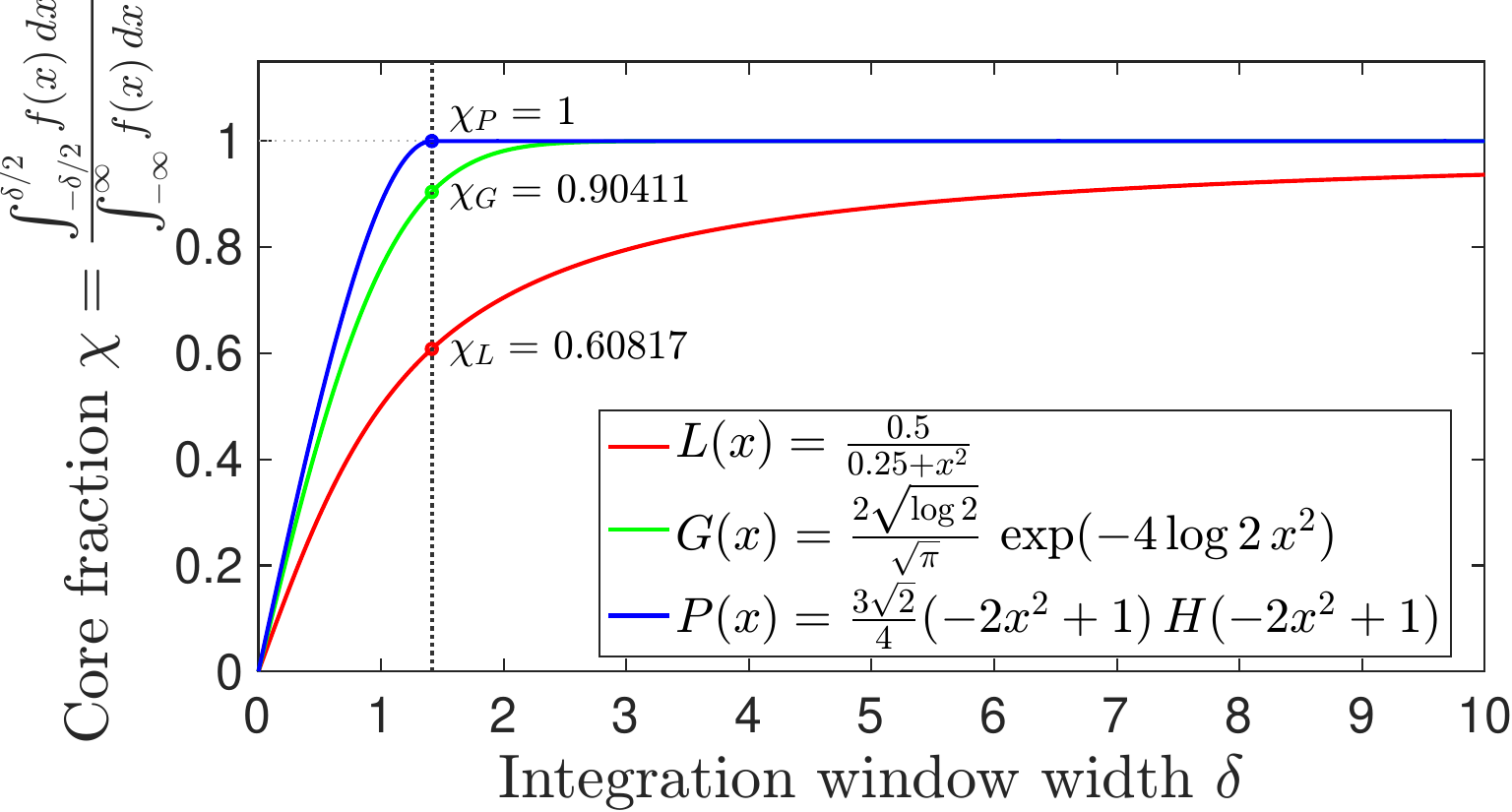}
\end{minipage}
\caption{\label{fig:CoreFraction}Function plots (top) and core fraction analysis (bottom) of a Lorentzian $L(x)$, a Gaussian $G(x)$, and a truncated parabola $P(x)$, all of which have FWHM and integral area~$1$. The truncated parabola has finite support and thus the least pronounced tails such that it reaches a core fraction of $\chi = 1$ fastest at an integration window width of $\delta=\sqrt{2}$. The tails of the Gaussian are less pronounced than those of the Lorentzian which results in a larger $\chi$ at the same value of $\delta$.}
\end{figure}

An intuitive way to quantitatively analyze the tails of line shape functions is given by the core fraction $\chi$ which corresponds to the ratio of the area obtained by integrating over a window of width $\delta$ centered at the extremum of the function and the total integral area. Functions with less pronounced tails approach a core fraction of $\chi=1$ faster. In Fig.~\ref{fig:CoreFraction}, a Lorentzian $L(x)$, a Gaussian $G(x)$, and a truncated parabola $P(x)$, each with a FWHM and integral area of $1$, are plotted and the corresponding core fractions $\chi$ are shown. It is clear, that the truncated parabola $P(x)$ is the fastest and – due to its finite support – only function to reach an exact core fraction of 1, whereas the Lorentzian and Gaussian approach $\chi = 1$ asymptotically. Comparing the core fractions at $\delta=\sqrt{2}$, where the parabola has $\chi_P=1$, the Lorentzian has the most pronounced tails resulting in $\chi_L \approx 0.61$ whereas the Gaussian tails are significantly less pronounced resulting in $\chi_L \approx 0.90$.

Furthermore, the behavior of parabolic lines in case of overlap is different to, e.g., Lorentzian and Gaussian lines and is discussed in Appendix~\ref{AppOverl}.

\subsection{Dispersion mode Lorentzians}\label{sec:DispLorSup}
\noindent
As was shown in FIG.~\ref{fig:MainCont}, the parametric continuous Newton transform (see section~\ref{sec:ch3}) can be applied to any combination of Lorentzian modes. However, from FIG.~\ref{fig:MainCont} it is clear that dispersion mode Lorentzian lines ($\varphi \mod \pi = \frac{\pi}{2}$) do not adopt parabolic line shapes under the introduced continuous parametric transform for which we cannot apply the findings from section~\ref{sec:AbsLorSup} in such cases. Here, $p = 0$ turns out to be the best case to obtain finite support narrow lines which approach the baseline rapidly away from the extrema.

\begin{figure}[htb!]
\centering
\includegraphics[width=0.41\textwidth]{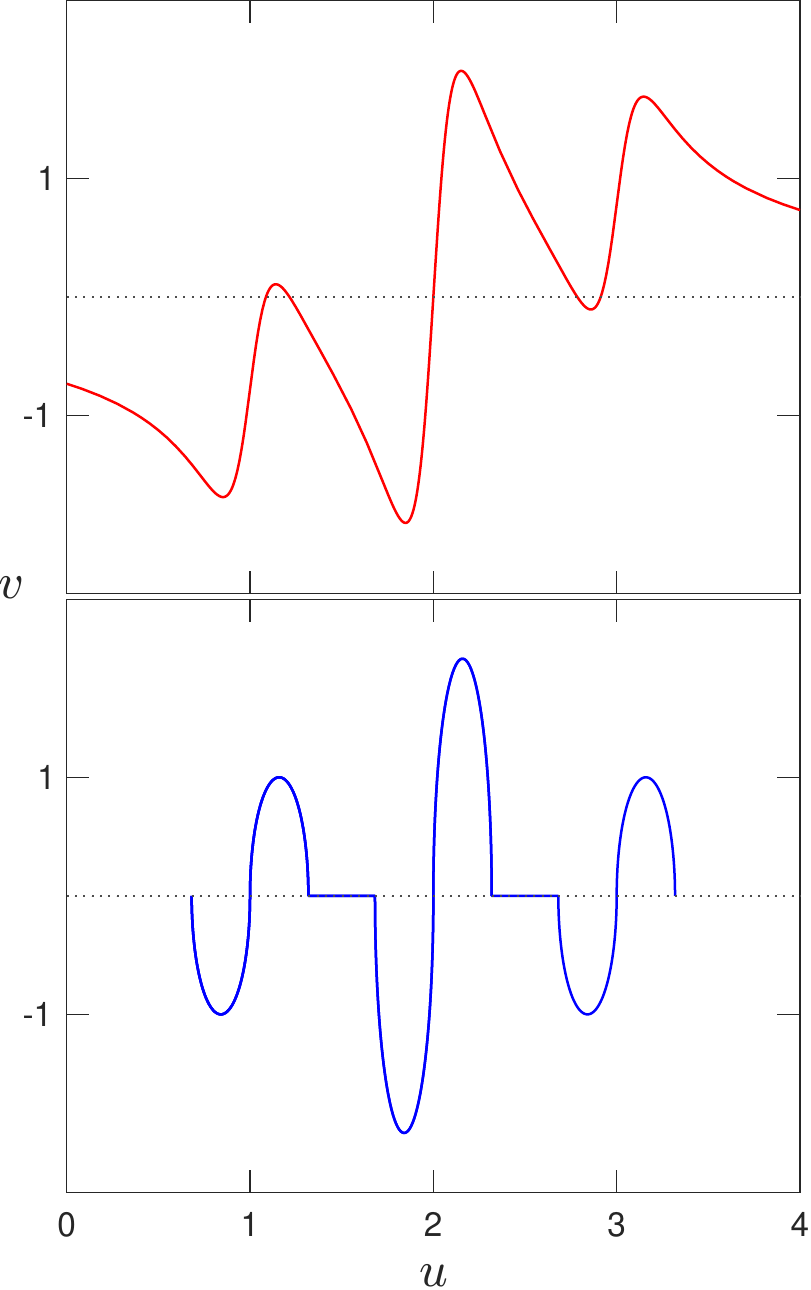}
\caption{\label{fig:DispSpec}Simulated triplet with a $90^\circ$ phase shift resulting in strongly overlapping pure dispersion mode Lorentzians (red). Applying a FWHM conserving parametric Newton transform variant from Eq.~\ref{eq:DispSupp} to every individual spectral line gives the ellipse spectrum (blue) without any overlap.}
\vspace{-.3cm}
\end{figure}

Indeed, using a transform parameter $p = 0$, a dispersion mode Lorentzian is transformed into corresponding semi-ellipse curves. Using the standard parametrization~A (cf. Tab.~\ref{tab:1} in Appendix~\ref{AppEll}), we obtain semi-ellipse lines with coordinates

\begin{eqnarray}
\label{eq:DispSupp}
u &&= x_{disp,FWHM} = \frac{k}{2\pi}\text{sgn}\left[\sin(\theta)\right]\left[\sin(\varphi)+\cos(\theta)\right]\nonumber\\
&&=\begin{cases*}
      {\frac{k}{2\pi}\left[\sin(\varphi)+\cos(\theta)\right]} & $\text{if } \theta = [0;\pi] $\\
      {-\frac{k}{2\pi}\left[\sin(\varphi)+\cos(\theta)\right]}& $\text{if } \theta = [\pi;2\pi]$ ,
    \end{cases*}
    \nonumber\\
      v &&= \frac{r}{k}\sin(\theta),
\end{eqnarray}

\noindent
in which case the FWHM of the pure absorption mode Lorentzian matches the horizontal semi-ellipse diameter at the baseline.
As can be seen in FIG.~\ref{fig:DispSpec}, this approach gives considerably sharper lines which do not show the pronounced tails of dispersion mode Lorentzians.

\section{A window function for parabolic line shapes}\label{sec:ch5}
\noindent
The previously discussed results are applicable for the efficient display of spectra but also motivate to study if potential experimental applications of this line shape are possible. As discussed below, we investigated whether it is possible to obtain parabolic spectral lines in experimental NMR spectra by designing a new window function for apodization purposes.

Transforming an absorption mode Lorentzian into a truncated parabola, as was discussed in section~\ref{sec:AbsLorSup}, gives narrow lines with significantly reduced support. This raises the question whether such line shapes can be reproduced in experimental settings.

Indeed, it is possible to design a window function that gives the desired line shape when applied in an apodization procedure similar to the well-established Lorentz-to-Gaussian transform \citep{Derome, Hoch} in NMR spectroscopy.

\begin{figure}[htb!]
\centering
\includegraphics[width=0.45\textwidth]{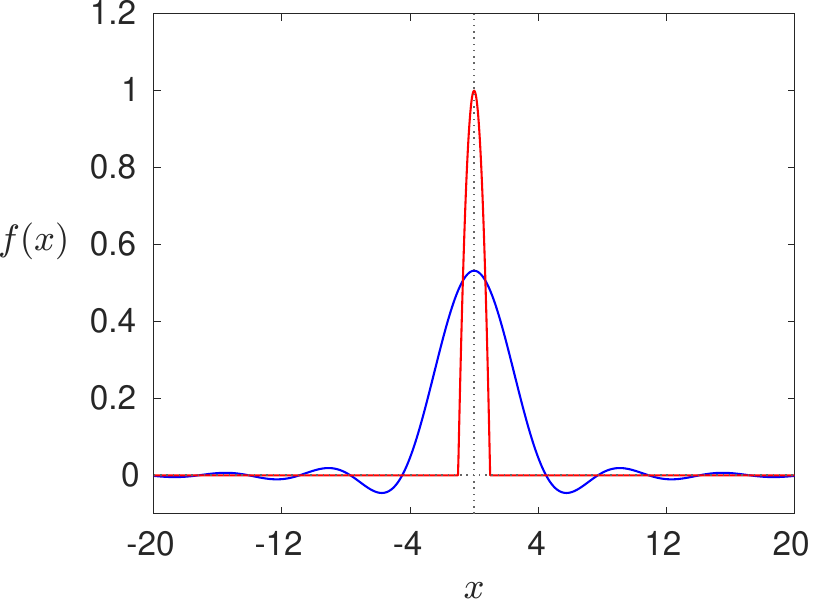}
\vspace{-.4cm}
\caption{\label{fig:WF}Truncated parabolic line shape function (red, Eq.~\ref{eq:FDFunc}) resulting from Fourier transforming the window function (blue, Eq.~\ref{eq:TDFunc}) for parametric values $s = 1$ and {FWHM~$=\sqrt{2}$}.}
\end{figure}

First, the relaxation-induced exponential decay of the measured oscillating (phase-corrected) time domain signal, which is usually referred to as free induction decay (FID)\citep{Bodenhausen, Keeler}, must be eliminated.
Second, the signal is multiplied with the new window function. This will cause all spectral lines in the Fourier (i.e, the frequency) domain to occur with parabolic shapes. Indeed, the window function can readily be derived from the parametric function describing the frequency domain line shape
%\begin{equation}
%\label{eq:FDFunc}
%S(\omega)=\big(s[-(a\omega)^2+1]\big)\frac{\text{sgn}(\big(s[-(a\omega)^2+1]\big))+1}{2}
%\end{equation}
%\begin{equation}
%\label{eq:FDFunc}
%\hspace{-.2cm}
%S(\omega)=s\bigg[-\bigg(\frac{\sqrt{2}\omega}{\text{FWHM}}\bigg)^2+1\bigg]\ H\bigg[-\bigg(\frac{\sqrt{2}\omega}{\text{FWHM}}\bigg)^2+1\bigg]
%\end{equation}
\begin{equation}
\label{eq:FDFunc}
\hspace{-.2cm}
S(\omega)=s\big(-(a\omega)^2+1\big)\ H\big(-(a\omega)^2+1\big)
\end{equation}
using the angular frequency $\omega$, an amplitude scaling factor $s$, the line-width scaling factor $a=(\sqrt{2}\pi\text{FWHM})^{-1}$ with the desired FWHM in Hertz, and the Heaviside function $H(\cdot)$.
We then proceed to calculate the window function $W(t)$ as the inverse Fourier transform of Eq.~\ref{eq:FDFunc} which results in
%\begin{equation}
%\label{eq:TDFunc}
%W(t)=F^{-1}_\omega [S(\omega)](t)=s\frac{a\:\big[8a\sin\big(\frac{t}{a}\big)-8t\cos\big(\frac{t}{a}\big)\big]}{2\sqrt{2\pi}t^3}.
%\end{equation}
%\begin{align}
%\label{eq:TDFunc}
%W(t)&=F^{-1}_\omega [S(\omega)](t)\\
%&=s\frac{\frac{4\sqrt{2}}{\text{FWHM}^2}\sin\big(\frac{t\:\text{FWHM}}{\sqrt{2}}\big)-\frac{4t}{\text{FWHM}}\cos\big(\frac{t\:\text{FWHM}}{\sqrt{2}}\big)}{\sqrt{\pi}t^3}.\nonumber
%\end{align}
\begin{align}
\label{eq:TDFunc}
W(t)&=F^{-1}_\omega [S(\omega)](t)\\
&=s\frac{2\sqrt{2}a}{\sqrt{\pi}}\ \frac{a\sin\big(\frac{t}{a}\big)-t\cos\big(\frac{t}{a}\big)}{t^3}.\nonumber
\end{align}
Both functions (Eqs.~\ref{eq:FDFunc} and \ref{eq:TDFunc}) are visualized in FIG.~\ref{fig:WF}.\newline
A simplified version of Eq.~\ref{eq:TDFunc} can be obtained by normalizing the window function to a maximum function value of 1 such that it only relies on the linewidth parameter FWHM:
%\begin{equation}
%\label{eq:TDaFunc}
%W_a(t)=3a\sqrt{\frac{\pi}{8}} \frac{a\:\big[8a\sin\big(\frac{t}{a}\big)-8t\cos\big(\frac{t}{a}\big)\big]}{2\sqrt{2\pi}t^3}.
%\end{equation}
%\begin{equation}
%\label{eq:TDaFunc}
%\widetilde{W}(t)=\frac{\frac{6\sqrt{2}}{\text{FWHM}^3}\sin\big(\frac{t\:\text{FWHM}}{\sqrt{2}}\big)-\frac{6t}{\text{FWHM}^2}\cos\big(\frac{t\:\text{FWHM}}{\sqrt{2}}\big)}{t^3}.
%\end{equation}
\begin{equation}
\label{eq:TDaFunc}
\widetilde{W}(t)=3a^2\frac{a\sin\big(\frac{t}{a}\big)-t\cos\big(\frac{t}{a}\big)}{t^3}.
\end{equation}
Both window functions (Eq.~\ref{eq:TDFunc} and Eq.~\ref{eq:TDaFunc}) are oscillating and possess infinitely many roots. In theory, this implies that a time domain signal apodized with one of the window functions, rather impractically, has to be infinitely long to yield perfect truncated parabolic line shapes after Fourier transform. 

\begin{figure}[htb!]
\centering
\includegraphics[width=0.46\textwidth]{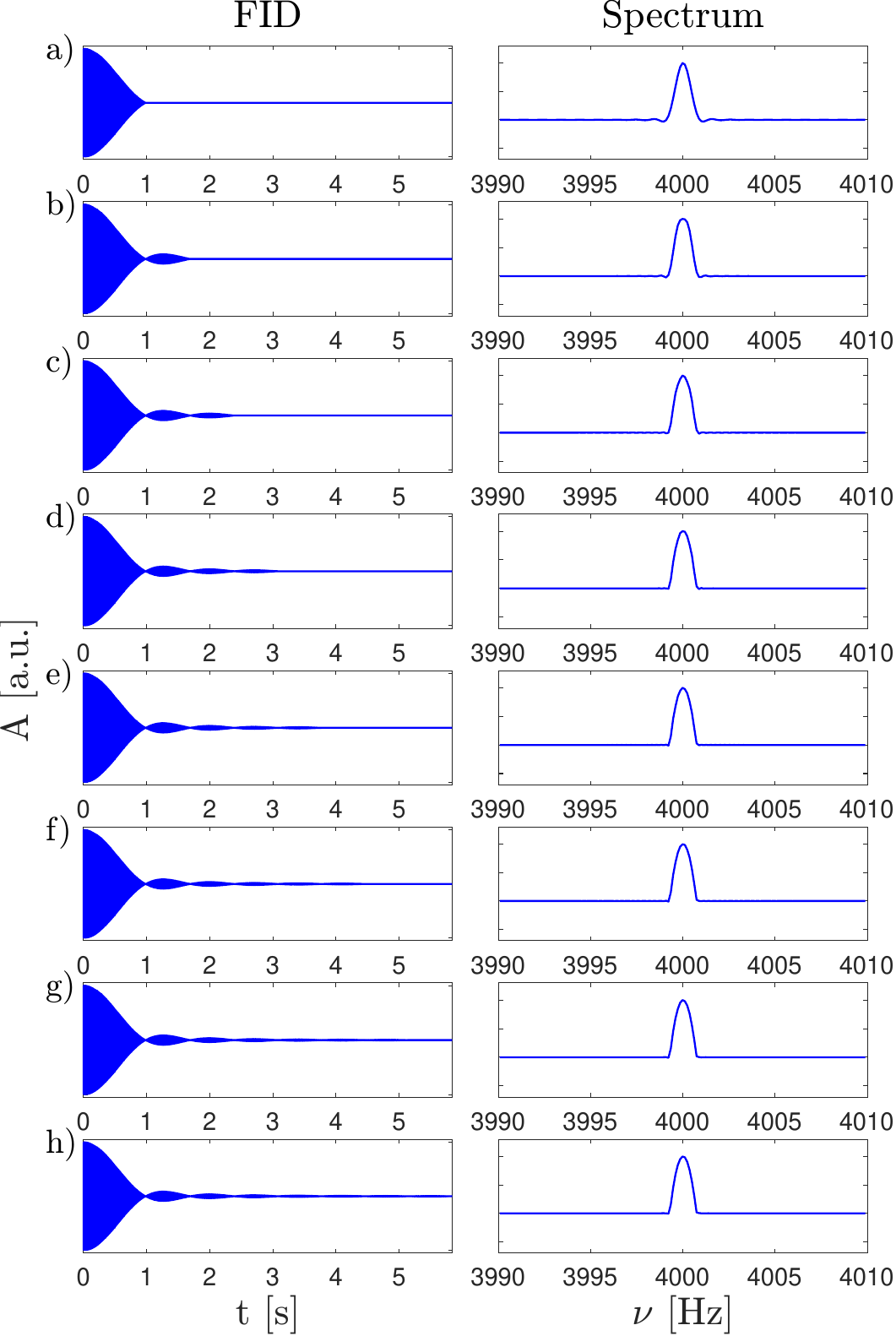}
\caption{\label{fig:Roots}Simulated FIDs and spectra using a single offset frequency $\nu = 4000$~Hz. The apodized time domain signal was truncated after the first (a) to the eighth root (h) of the applied window function (Eq.~\ref{eq:TDaFunc}, $\text{FWHM} = 1\:\text{Hz}$) and zero-filled to achieve identical spectral resolutions. The desired parabolic line shapes can already be obtained by truncation after the third root and do not change considerably for any of the later truncations.}
\vspace{-.7cm}
\end{figure}

\begin{figure}[htb!]
\centering
\includegraphics[width=0.43\textwidth]{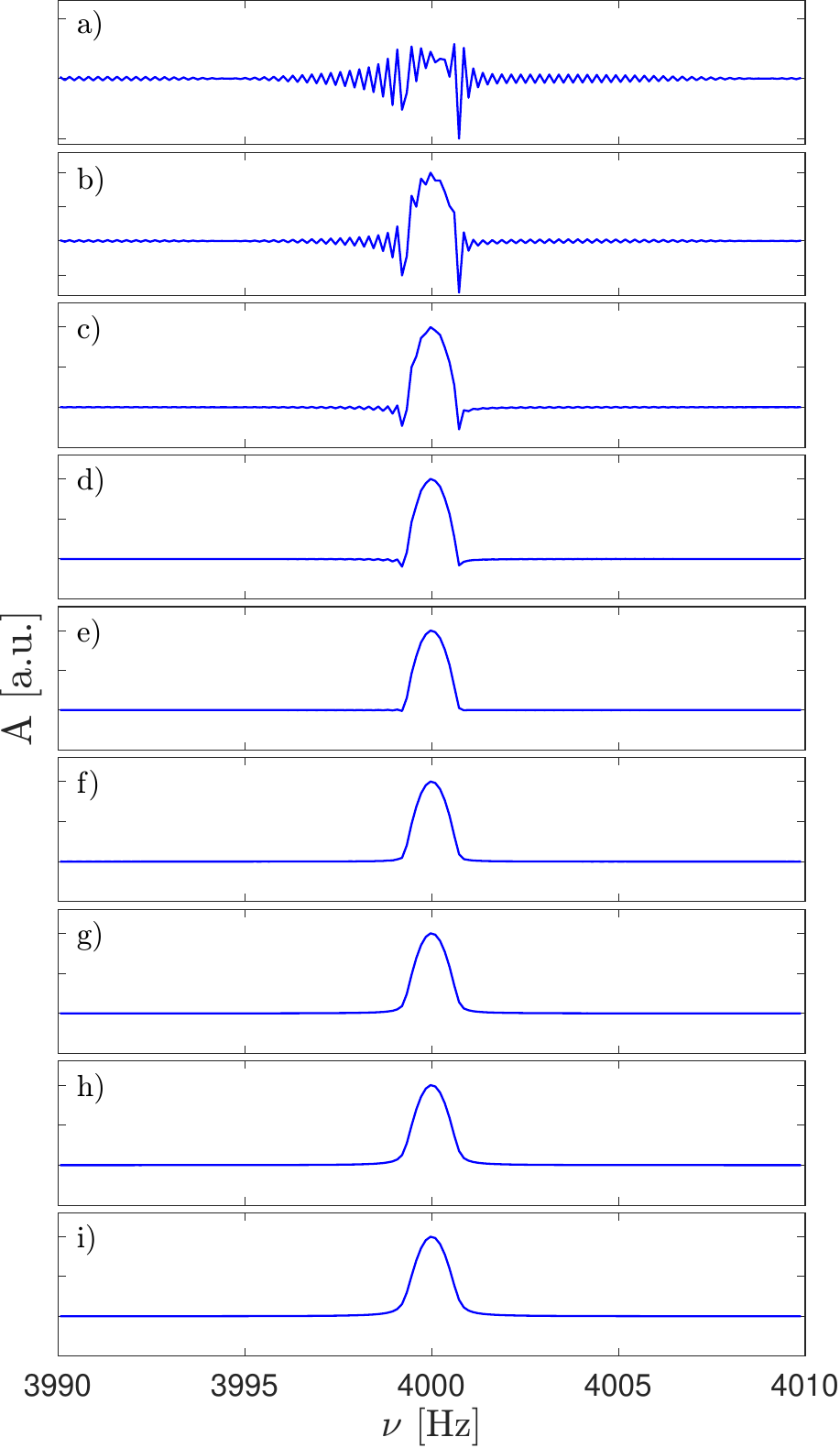}
\caption{\label{fig:Comp}Simulated spectra using a single offset frequency $\nu = 4000$~Hz. The original time signal decay corresponding to a damping constant $k = \frac{1}{T_2} = \big(\frac{1}{\pi} \text{s}\big)^{-1} \approx 3.14$~Hz was compensated by multiplying the signal with an exponential function $f_{\text{comp}} = \exp(k_{\text{comp}}t)$ using compensation constants $k_{\text{comp}} = T_{2,\text{comp}}^{-1}$ with $T_{2,\text{comp}}$-values ranging from $\frac{2}{3}T_2$ (a) to $\frac{4}{3}T_2$ (i) in steps of $\frac{1}{12}T_2$. Case e) thus corresponds to a perfectly compensated decay. All FIDs were apodized to give parabolic lines in the frequency domain with a line width of $\text{FWHM} = 1\:\text{Hz}$. The resulting lines possess Lorentzian contributions (undercompensation, cases f) to i)) or become distorted in overcompensation cases ( examples a to d).}
\end{figure}

\begin{figure}[ht!]
\centering
\includegraphics[width=0.5\textwidth]{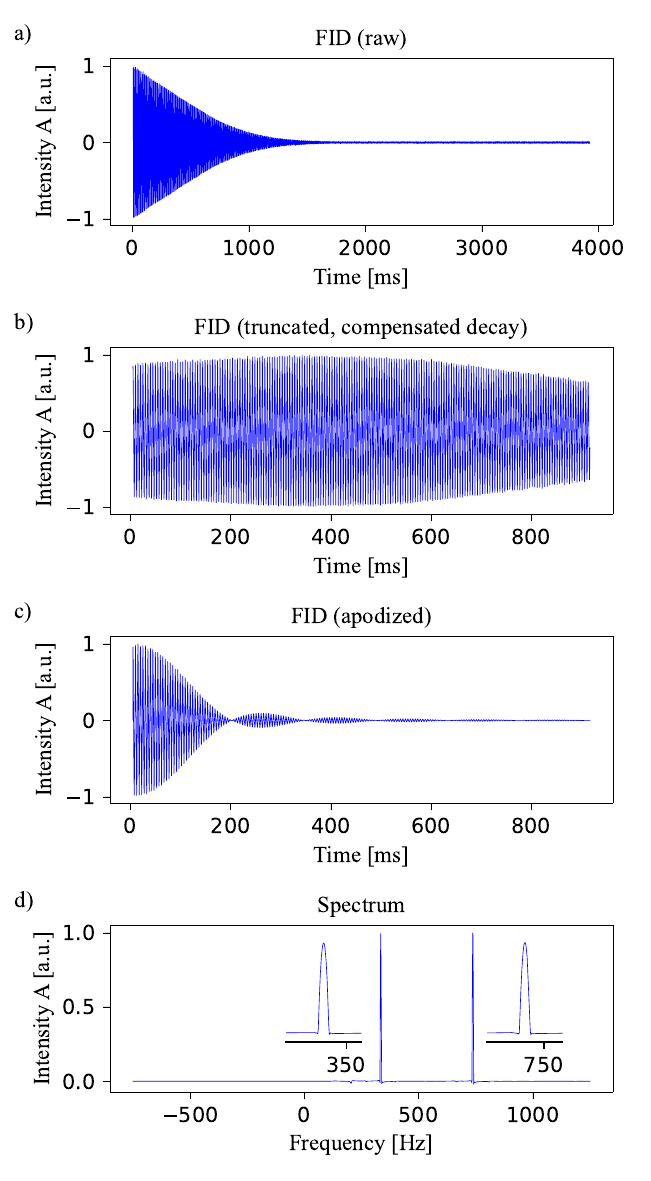}
\caption{\label{fig:MASpec}NMR data of methyl acetate recorded at a spectrometer frequency of  $\nu = 250$~MHz. (a) the raw FID is first phase corrected and truncated. (b) then, the naturally occuring exponential decay of the signal is compensated. (c) The resulting oscillating signal is multiplied  with the window function (Eq.~\ref{eq:TDaFunc}). (d) after Fourier transform, both singlets in the frequency domain possess the desired parabolic line shape . Both lines are additionally displayed in 50 Hz-wide close-up views.}
\end{figure}

The introduced window function is oscillating around the baseline. This suggests to truncate the time signal at one of the function roots $\zeta_a$ which are given by the solutions of the equation $\text{tanc}(x)-1 = 0$ multiplied by the apodization parameter $a$, where $\text{tanc}(x)=\frac{\tan(x)}{x}$ is the cardinal tangent function. Doing so will minimize the occurrence of so-called sinc wiggling artifacts in the Fourier domain which would occur as the Fourier transform of the step function envelope curve that is obtained when truncating the time-domain signal at other positions.

FIG.~\ref{fig:Roots} shows a simulated time domain signal, using an offset frequency of $\nu = 4000$~Hz that was apodized by using the introduced window function Eq.~\ref{eq:TDaFunc}. The signal was then truncated at the roots of the window function and Fourier transformed. The resulting spectra are also shown in FIG.~\ref{fig:Roots}. It is clear, that a parabolic line shape is already closely approached by truncating the signal at the second or third root of the window function.

We further examined the robustness of the apodization approach with respect to imperfect compensations of naturally occurring time signal exponential decays before multiplying the signal with the introduced window function. Indeed, this is important for experimental applications in which multiple differing damping constants may appear for different residues of investigated molecules or materials, as in such cases not all signal decays can be compensated exactly.

Hence, we simulated the resulting line shapes for over- and undercompensated exponential decays of the same signal. As it is clear in Fig.~\ref{fig:Comp}, undercompensation leaves a residual exponential decay that introduces Lorentzian contributions to the parabolic line shape and overcompensation leads to a distortion of the line shape that ranges from small knees at the baseline to severe wiggling. However, the simulations reveal a satisfactory robustness of the proposed apodization method with respect to inaccurate compensation. Hence, the apodization only becomes inapplicable in cases of severe overcompensation.  

We further confirmed the functionality of our approach by applying it to experimentally measured data of the chemical compound methyl acetate. As can be seen in FIG.~\ref{fig:MASpec}, it is clear that the spectral lines are indeed transformed into the desired parabolic shapes.\newline
Additional simulations of FIDs including noise are provided in appendix App.~\ref{AppNoise}.

\section{Discussion}\label{sec:ch6}
\noindent
The Newton hyperbolism transform of curves provides a unique conceptual approach to connect key concepts of modern Fourier transform spectroscopy with Bloch vectors \citep{Bloch, Feynman} and related phase-space representations of underlying quantum systems. We demonstrated how phase-sensitive FT spectra can be derived starting from a simple Bloch picture of spectroscopically relevant single transitions using the introduced transform. 

Hence, a direct geometric connection between spectroscopic observables and theoretical representation approaches could be established.

The presented transform is surprisingly simple and does only require elementary geometry and function analysis. It is not only easy to perform but it is also exceptionally simple to grasp. Thus, the approach may also be well-suited for educational purposes.

We discovered that finite support parabolic line shapes represent a special intermediate case of the continuous parametric Newton transform between ellipses and corresponding Lorentzians. This served as a motivation for the development of a new apodization approach similar to the prominent Lorentz-to-Gaussian transform \citep{Derome, Hoch} which allowed to replicate parabolic line shapes in experimental nuclear magnetic resonance spectra to a good approximation. We also introduced parametrizations of the dedicated window functions which allow to manipulate the width of these parabolic lines. This apodization technique may well serve as a starting point for future work to study potential practical advantages, e.g., in multidimensional spectroscopy and automatized peak picking.

\section{Conclusion}\label{sec:ch7}
\noindent
In this work, we generalized the concept of the Newton hyperbolism transform of curves to a group of parametric continuous transforms that allows to convert ellipses into Lorentzian lines and vice versa. With this approach, we were able to establish a direct geometric connection between phase sensitive FT spectra and the Bloch picture or equivalent phase-space representations describing the underlying detectable transitions. To do so, only little computational effort and basic geometric operations are required whilst Fourier transform can be completely avoided.
% We implemented a spectrum simulation using the introduced findings in the spin simulation software \textit{SpinDrops}.

Furthermore, parametrization of the presented transform allowed to identify special cases of line shapes with finite support which served as a main motivation for the development of a new window function for parabolic line shapes in Fourier transform spectroscopy. We were also able to apply the latter successfully in a proof-of-concept NMR apodization procedure. The application of this technique in multi-dimensional spectroscopy is also promising.
 
\begin{acknowledgments}
\noindent
D.H. and S.J.G. acknowledges support from the \textit{Verband der chemischen Industrie e.V (VCI)} and \textit{TUM EDU} (grant no. 3170573). The authors acknowledge funding from the Digital Europe Programme for the project DigiQ: Digitally Enhanced Quantum Technology Master (DigiQ) under grant agreement ID 101084035.

We thank F. vom Ende and E. Malvetti for the helpful mathematical discussions and R. Marx and S. Lohmaier for providing the utilized NMR data. Research and experiments were performed at the Bavarian NMR Center (BNMRZ) at the Technical University Munich.
\end{acknowledgments}

\newpage
\appendix
\section{Parameterized ellipses and their hyperbolism transform in spectroscopic applications}\label{AppEll} 
\noindent
Extending the procedure presented in section~\ref{sec:ch1}, it is possible to reparametrize the circles obtained from the corresponding Bloch vector or phase-space representation to give physically scaled ellipses. This allows to reproduce important properties of Lorentzian spectral lines such as the amplitude, FWHM, or integral areas in corresponding ellipse representations.

Indeed, such ellipses result from scaling a circle similar to what was shown for Lorentzian lines in \textit{Step~6} of section~\ref{sec:ch1}. However, this scaling is applied before the actual Newton transform (i.e., before \textit{Step~4}) and makes \textit{Step~6} no longer necessary. This may be useful if cartoon-like spectrum approximations are desired that remain closely related to the original Bloch or phase-space pictures and can be obtained in a few steps. Four exemplary parametrizations A–D are presented in Tab.~\ref{tab:1}.

Ellipse parametrization~A is derived by analyzing the absorption mode Lorentzian function (Eq.~\ref{eq:AbsLorPar}) which has a FWHM~$= \frac{k}{\pi}$. X-coordinates of points on the ellipse are then scaled with a factor $\frac{k}{2\pi}$ such that the ellipse width matches the corresponding Lorentzian FWHM (see FIG.~\ref{fig:MainCont}). In the main text, we use this parametrization for our analyses.\newline
Parametrization~B is simpler and only reproduces the Lorentzian amplitude in the corresponding ellipse.

Parametrizations~C and D are chosen such that the ellipse area is identical to the corresponding pure absorption mode Lorentzian integral area which is given by
\begin{equation}
\int A(\nu)\:d\nu = r\frac{\arctan(2\pi\frac{\nu}{k})}{\pi}\nonumber.
\end{equation}
Knowing that $\lim\limits_{\nu\rightarrow\pm\infty} \arctan\big(2\pi\frac{\nu}{k}\big) = \pm \frac{\pi}{2}$, it directly follows that
\begin{equation}
\int_{-\infty}^\infty A(\nu)\:d\nu = r\frac{\arctan\big(2\pi\frac{\nu}{k}\big)}{\pi}\:\Bigg|^\infty_{-\infty}=\frac{r}{2}-\bigg(-\frac{r}{2}\bigg)=r.
\end{equation}
The ellipse semi-minor axis $b$ (corresponding to the ellipse width) can then be calculated based on the ellipse area $\mu_E$ which is given by
\begin{equation}
\mu_E = \pi ab = r
\end{equation}
using the semi-major axis $a$, e.g., $a = \frac{r}{k}$ (parametrization~C) or $a = \frac{2r}{k}$ (parametrization~D) which gives $b = \frac{k}{\pi}$ or $b = \frac{k}{2\pi}$, respectively. Parametrization~D also conserves the absorptive Lorentzian FWHM.

Continuous parametric Newton transform expressions and properties of all introduced ellipse parametrizations are summarized in Tab.~\ref{tab:2}.

\begin{figure}[htb!]
\centering
\includegraphics[width=0.47\textwidth]{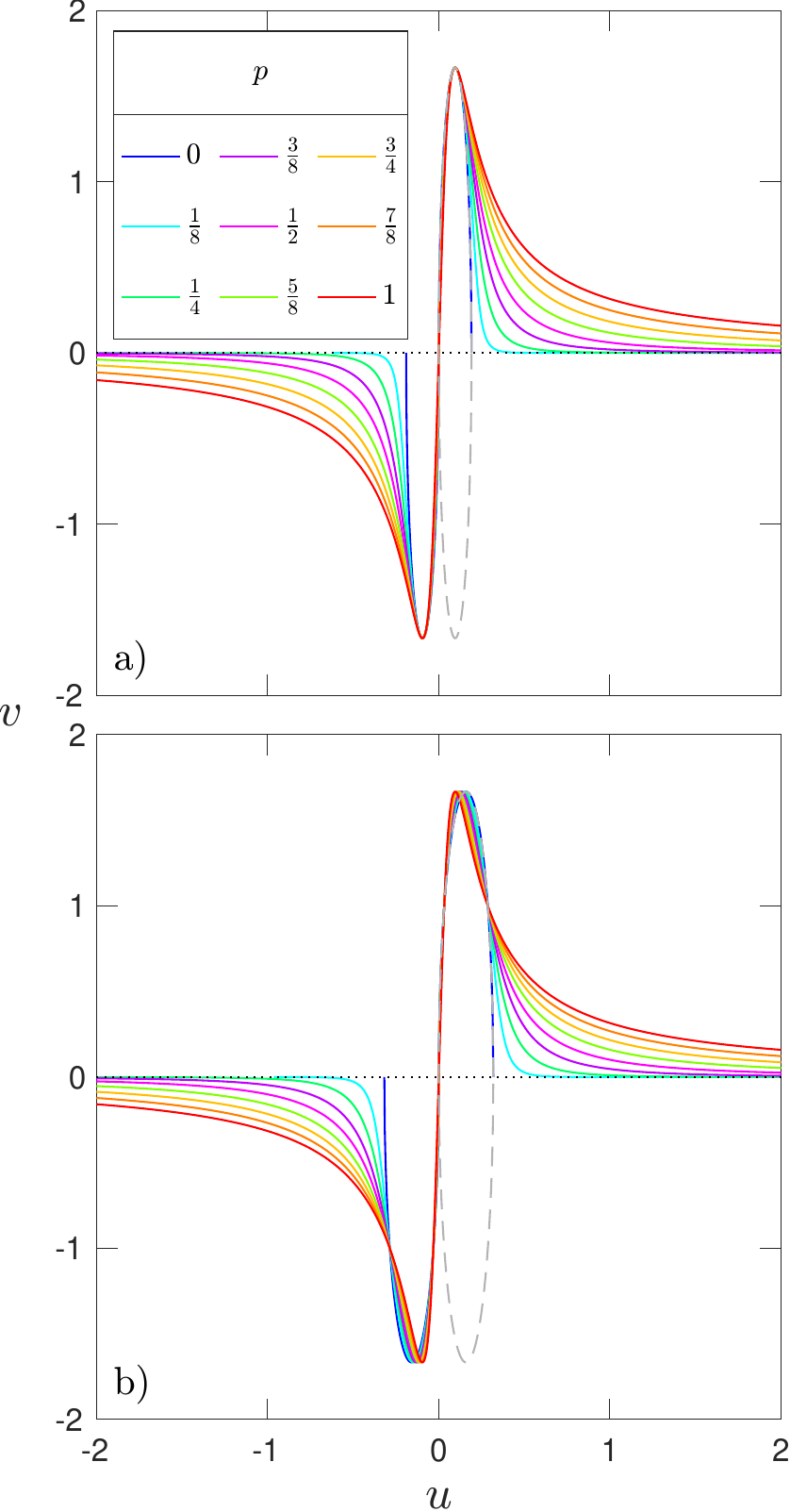}
\caption{\label{fig:ContDisp}Continuous transforms of a pure dispersion mode Lorentzian ($\varphi = \frac{\pi}{2},\:r = 1,\:k=0.6$) using parametrizations~A (a) and B (b), see TAB.~\ref{tab:1} and TAB.~\ref{tab:2} for details). In case (b), the $u$-coordinates of the extrema of parameterized curves corresponding to $p\neq1$ do not coincide with those of the Lorentzian. Conversely, both coordinates of the extrema always coincide with those of the Lorentzian for transform (a). The original ellipse is indicated by a dashed grey line and the baseline is marked by a dotted horizontal line in both cases.}
\end{figure}

For instance, parametrization~A is used in FIG.~\ref{fig:MainCont} of section~\ref{sec:ch3} to visualize the new parametric continuous Newton transform for different Lorentzian modes.
FIG.~\ref{fig:ContDisp} additionally shows a comparison of dispersion Lorentzian curves that result from transforming ellipses using parametrizations~A and B. It is clear that in this case transformations using parametrization~A leave the coordinates of extrema unchanged which is not the case for parametrization~B.

\vfill

\begin{table*}[htb!]
\centering
\caption{\label{tab:1}Parameterized ellipses that preserve distinct properties of spectral Lorentzian lines can be obtained by scaling the corresponding projected phase-space representation of a single-transition. Here, $r$ denotes the circle radius (see Eq.~\ref{eq:radius} in section~\ref{sec:ch1}) and $k$ is the damping constant associated to the observed transition which can be retrieved from the exponential decay of a corresponding measured time-domain signal. $\varphi$ denotes the phase of the transition. We introduce four exemplary ellipse parametrizations which are further characterized with respect to their properties and the corresponding continuous Newton transform in TAB.~\ref{tab:2}.}
\vspace{.3cm}
\begin{tabular}{m{0.15\textwidth}<{\centering} m{0.4\textwidth}<{\centering} p{0.3\textwidth}<{\centering}}
\hline\hline\\[-2.5pt]
Parametrization & Parametric expression $P_E = (x_E,y_E) =$ & Ellipse equation\\[7.5pt] \hline \\[-2.5pt]
A & $\bigg(\frac{k}{2\pi}\left[\sin(\varphi)+\cos(\theta)\right],\frac{r}{k}\left[\cos(\varphi)+\sin(\theta)\right]\bigg)$ & $\frac{\big(x-\frac{k}{2\pi}\sin(\varphi)\big)^2}{\big(\frac{k}{2\pi}\big)^2}+\frac{\big(y-\frac{r}{k}\cos(\varphi)\big)^2}{\big(\frac{r}{k}\big)^2} = 1$ \\[15pt]
B & $\bigg(\frac{r}{2\pi}\left[\sin(\varphi)+\cos(\theta)\right],\frac{r}{k}\left[\cos(\varphi)+\sin(\theta)\right]\bigg)$ & $\frac{\big(x-\frac{r}{2\pi}\sin(\varphi)\big)^2}{\big(\frac{r}{2\pi}\big)^2}+\frac{\big(y-\frac{r}{k}\cos(\varphi)\big)^2}{\big(\frac{r}{k}\big)^2} = 1$  \\[15pt]
C & $\bigg(\frac{k}{\pi}\left[\sin(\varphi)+\cos(\theta)\right],\frac{r}{k}\left[\cos(\varphi)+\sin(\theta)\right]\bigg)$ & $\frac{\big(x-\frac{k}{\pi}\sin(\varphi)\big)^2}{\big(\frac{k}{\pi}\big)^2}+\frac{\big(y-\frac{r}{k}\cos(\varphi)\big)^2}{\big(\frac{r}{k}\big)^2} = 1$ \\[15pt]
D & $\bigg(\frac{k}{2\pi}\left[\sin(\varphi)+\cos(\theta)\right],\frac{2r}{k}\left[\cos(\varphi)+\sin(\theta)\right]\bigg)$ & $\frac{\big(x-\frac{k}{2\pi}\sin(\varphi)\big)^2}{\big(\frac{k}{2\pi}\big)^2}+\frac{\big(y-\frac{2r}{k}\cos(\varphi)\big)^2}{\big(\frac{2r}{k}\big)^2} = 1$ \\[15pt] \hline\hline
\end{tabular}
\vspace{.0cm}
\caption{\label{tab:2}Continuous parametric transform expressions of the ellipses described in Tab.~\ref{tab:1}. Using the parameter $p$ it is possible to continuously transform any point $P_p$ on a parameterized ellipse ($p = 0$) to the same Lorentzian spectral line ($p = 1$, see Eq.~\ref{eq:LorP}), respectively. Here, $\zeta=\text{sgn}(y_p) = \text{sgn}{(\cos{\varphi}+\sin{\theta}})$. $\bullet$, $\circ$, and $\times$ symbols indicate which of the denoted properties do fully, partially (i.e., only for particular values of $p$), or not apply to a parametrized ellipse, respectively. FWHM$_{abs}$ refers to the full width at half maximum of the absorption mode Lorentzian. Note that we use parametrization~A for visualizations in the main text.}
\vspace{.3cm}
\begin{tabular}{p{0.13\textwidth}<{\centering} p{0.45\textwidth}<{\centering} m{0.1\textwidth}<{\centering} m{0.1\textwidth}<{\centering} m{0.1\textwidth}<{\centering}}
\hline\hline\\[-2.5pt]
Parametrization & Continuous transform $P_p = (x_p,y_p) =$ & Amplitude & FWHM$_{abs}$ & Integral \\[7.5pt] \hline \\[5pt]
A & $\Bigg(\frac{\zeta\frac{r^pk^{1-p}}{2\pi}\left[\sin(\varphi)+\cos(\theta)\right]}{\big|\frac{r}{k}\left[\cos(\varphi)+\sin(\theta)\right]\big|^p},\frac{r}{k}\left[\cos(\varphi)+\sin(\theta)\right]\Bigg)$ & \scalebox{2}{$\bullet$} & \scalebox{2}{$\bullet$} & \scalebox{1.5}{$\times$} \\[15pt]
B & $\Bigg(\frac{\zeta\frac{r}{2\pi}\left[\sin(\varphi)+\cos(\theta)\right]}{\big|\frac{r}{k}\left[\cos(\varphi)+\sin(\theta)\right]\big|^p},\frac{r}{k}\left[\cos(\varphi)+\sin(\theta)\right]\Bigg)$ & \scalebox{2}{$\bullet$} & \scalebox{1.5}{$\times$} & \scalebox{1.5}{$\times$} \\[15pt]
C & $\Bigg(\frac{\zeta\frac{r^pk^{1-p}}{2^p\pi}\left[\sin(\varphi)+\cos(\theta)\right]}{\big|\frac{r}{k}\left[\cos(\varphi)+\sin(\theta)\right]\big|^p},\frac{r}{k}\left[\cos(\varphi)+\sin(\theta)\right]\Bigg)$ & \scalebox{2}{$\bullet$} & \scalebox{1.5}{$\times$} & \scalebox{2}{$\circ$}\\[15pt]
D & $\Bigg(\frac{\zeta\frac{r^pk^{1-p}}{2\pi}\left[\sin(\varphi)+\cos(\theta)\right]}{\big|\frac{r}{k}\left[\cos(\varphi)+\sin(\theta)\right]\big|^p},\frac{2^{1-p}r}{k}\left[\cos(\varphi)+\sin(\theta)\right]\Bigg)$ & \scalebox{2}{$\bullet$} & \scalebox{2}{$\bullet$} & \scalebox{2}{$\circ$}\\[15pt] \hline\hline
\end{tabular}
\end{table*}

\noindent
Note that even though the ellipses differ, the full transform case ($p = 1$) always yields the same Lorentzian which can be expressed parametrically as
\begin{equation}
\label{eq:LorP}
P_{L} = \bigg(\frac{k}{2\pi}\frac{\left[\sin(\varphi)+\cos(\theta)\right]}{\left[\cos(\varphi)+\sin(\theta)\right]},\frac{r}{k}\left[\cos(\varphi)+\sin(\theta)\right]\bigg).
\end{equation}

Furthermore, parametrizations~C and D only reproduce the Lorentzian integral area in case of the ellipse ($p=0$) the Lorentzian itself ($p=1$) but not for any other value of $p$ (see TAB.~\ref{tab:2}).

\section{Expanding the transform parameter range}\label{AppE}
\noindent
The continuous parametric transform introduced in section~\ref{sec:ch3} only uses transform parameter values in the bounded interval $[0;1]$. However, $p$ is in general not limited to the denoted range and can be applied using any other arbitrary value.
Here, we briefly discuss the effect of transform parameters $p < 0$ and $p > 1$.

\begin{figure}[H]
\centering
\hspace{-.4cm}
\includegraphics[width=0.45\textwidth]{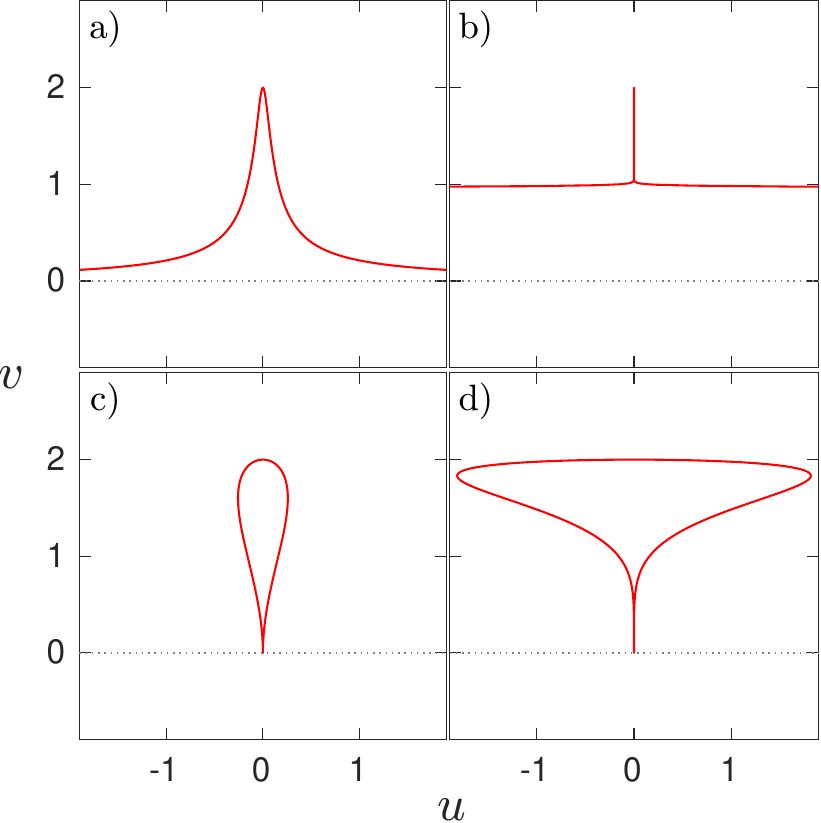}
\caption{\label{fig:OthParms}Continuous transform of a pure absorption mode Lorentzian ($\varphi = 0,\:r = 1,\:k=0.6$) using transform parameters (a) $p = 1.5$ , (b) $100$, (c) $-1.5$ and (d) $-5$. Values $p>1$ cause transformations towards delta-function-like line shapes on a pedestal. Conversely, transform parameters $p<0$ yield drop- or mushroom-like shapes.}
\end{figure}

As can be seen in FIG.~\ref{fig:OthParms}, for $p > 1$, the curve forms a pedestal for increasing values of $p$ which approaches values of $\frac{r}{k}$ in the $v$-dimension while the remaining peak converges to a delta-function-like shape.

%Indeed, this behavior may serve as a basis for future research, as herewith, it is possible to narrow spectral lines to a maximum degree.
For parametric values $p < 0$, the resulting transformed lines resemble upside-down raindrop shapes.~While transform parameters with large magnitude, i.e., $p\ll0$, cause broadened mushroom-like curves, parametric values close to zero yield narrow drop shapes with a flattened round top and a pointed base which in fact slightly resemble arrows.
%which could be, independent of this work, be interesting for new visual representations of quantum dynamics
Indeed, special curves obtained for parametric values $p = -1$ are commonly referred to as \textit{piriform quartics} {\citep{Wallis, Longchamps}} which, in the frame of this work, can be expressed parametrically as
\begin{equation}
P_{pq} = \big(\eta_x \cos(\theta) [1+\sin(\theta)], \eta_y [1+\sin(\theta)]\big),
\end{equation}
where $\eta_{x,y}$ denote coordinate scalings that depend on the chosen parametrization of the original circle or ellipse and $\theta$ is the polar angle.

\section{Overlapping parabolic line shapes}\label{AppOverl}
\noindent
While parabolic line shapes have finite support, they show a special behavior in case of overlapping lines that may appear rather cumbersome at first but can in fact be advantageous compared to the overlap properties of Lorentzian or Gaussian lines.

Indeed, the sum curves of overlapping parabolic lines are also parabolic. This arises from the mathematical properties of truncated quadratic funcions. However, if the extremal distance of two overlapping truncated parabolic spectral lines $\Delta \nu \geq \frac{\sqrt{2}\text{FWHM}}{2}$, the occurrence of additional lines leaves the positioning of all original maxima or minima unaltered which is not the case, e.g., for Lorentzians. Note that for cases in which $\Delta \nu < \frac{\sqrt{2}\text{FWHM}}{2}$ both parabolic lines blend into a single line such that the original extrema become invisible. 

\begin{figure}[t!]
\centering
\includegraphics[width=0.48\textwidth]{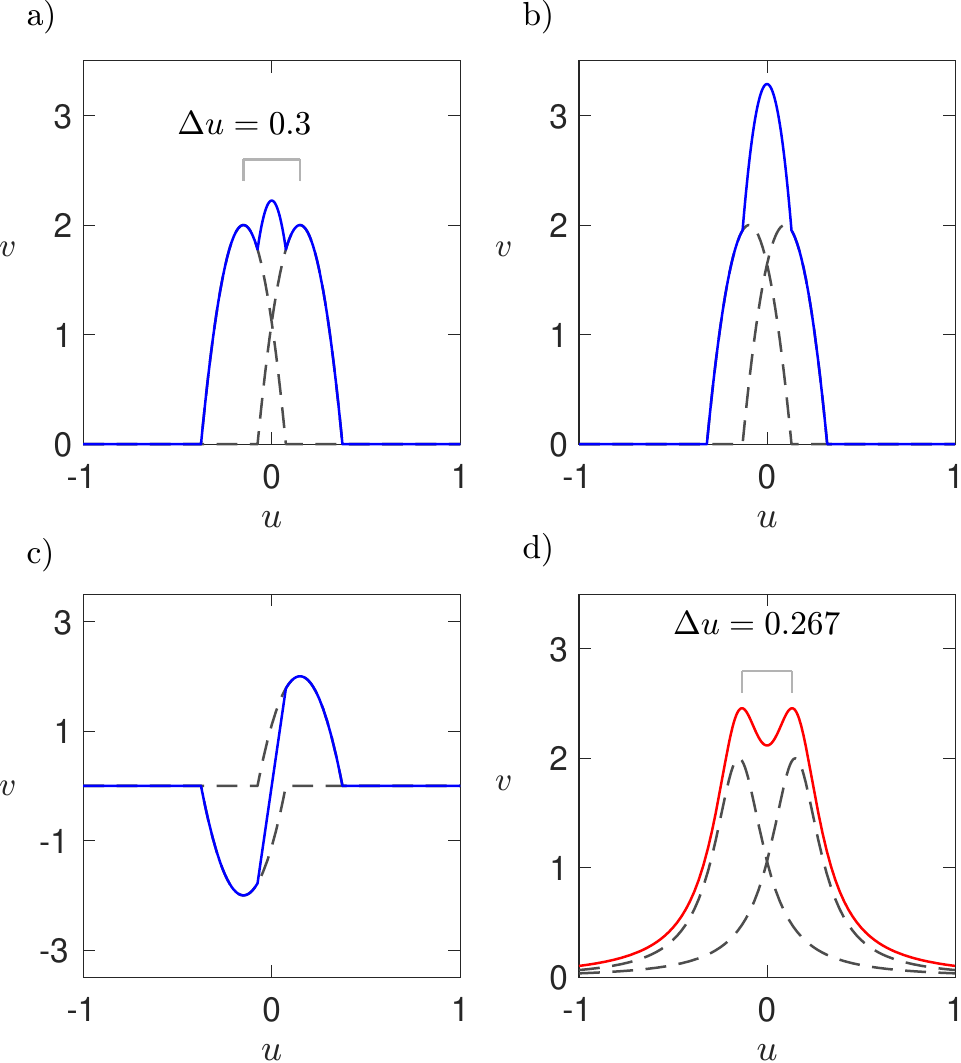}
\caption{\label{fig:ParabOver}Simulated overlapping doublet of truncated parabolas (a$-$c, with $r = 1$ and $k = 1$) and (d) of Lorentzians. The overlap causes the overall curve to adopt a pseudo-triplet shape (a) as $\Delta u = 0.3 > \frac{\sqrt{2}\text{FWHM}}{2}$ with $\text{FWHM} = \frac{1}{\pi}$. Both extrema of the original lines remain separated by $\Delta u$. In case b), the distance between the individual lines is $\Delta u = 0.19 < \frac{\sqrt{2}\text{FWHM}}{2}$ which yields a single distorted line. Overlapping parabolic lines with opposite phases leads to a cancellation of quadratic components, leaving only linear components in the overlapping regions (c). Overlapping Lorentzians with identical FWHM (d) yields a smooth curve but the extrema are shifted, i.e., they converge in the $u$-dimension such that a distance between maxima of $\Delta u \approx 0.267$ instead of the actually chosen distance $\Delta u = 0.3$ can be measured. All individual lines and their overlapping regions are indicated by grey dashed lines.}
\end{figure}

It should be stressed that this even holds true if two overlapping lines differ in phase by $180^\circ$. This is, e.g., the case for anti-phase magnetization in NMR \citep{Keeler}. However, for such combinations, the overlap does not generate additional parabolic lines but leads to a cancellation of quadratic contributions leaving only linear components, i.e., a straight line that connects both non-overlapping regions.

The discussed properties imply that when spectral multiplets are observed which arise due to coupling, e.g., scalar coupling in NMR, it is always possible to measure the physically correct coupling constant $J$ if $\Delta\nu = J \geq  \frac{\sqrt{2}\text{FWHM}}{2}$.

In contrast, the significant tails of Lorentzian or Gaussian lines cause their extrema to shift in case of overlaps with adjacent lines for which one only measures scaled coupling constants $\tilde{J}$.

Graphical examples of all described cases are shown in FIG.~\ref{fig:ParabOver}. Here, the partial overlap of two parabolic lines induces an overall pseudo-triplet-like curve appearance whereas increased overlaps yields a single maximum with piecewise-parabolic shape. Overlapping truncated parabolas with opposite signs cause a cancellation of quadratic components that is readily identifiable by a residual straight line. Identically parameterized Lorentzian lines experience a scaling of the extremal distance upon overlap. 

Note that even in cases where the original extrema of truncated parabolic lines lie hidden within overlapping regions, the mathematical properties of this line shape allow to apply straightforward curve fitting to determine the original extremal positions. Moreover, the derivative of the sum curve of overlapping truncated parabolas is not continuous at the boundary points of overlapping regions which may facilitate the deconvolution of these overlapping signals.

\section{Performance of parabolic line shape apodizations in case of noise-containing signals}\label{AppNoise}
\noindent
Although the simulations provided in section~\ref{sec:ch5} suggest that the proposed parabolic line shape apodization works as intended, they do not consider noise which is unavoidable in experimental settings. Similarly, a highly concentrated sample for the conducted experiment (see FIG.~\ref{fig:MASpec}) was used which allowed to obtain a high signal-to-noise ratio. In the following, we hence provide additional simulations to assess the robustness of our apodization approach in presence of strong noise.

Note that all simulations were carried out using normally distributed white noise profiles for the real and imaginary parts of a simulated single-frequency oscillation only. However, one may encounter other types of noise, e.g., Brownian or pink noise, in experimental applications.

Noise profiles were generated computationally and added to a clean exponentially decaying oscillating time-domain-signal. The noise power $P_{\text{noise}}$ was chosen to yield approximate signal-to-noise ratios ${\text{S/N} = \frac{P_{\text{sig}}}{P_{\text{noise}}} \approx 10}$ in the frequency domain. Be aware, that this corresponds to relatively strong noise in the time domain. The window function parameters (Eq.~\ref{eq:TDaFunc}) were set to give an FWHM of 1~Hz for all simulations.

\begin{figure}[H]
\centering
\includegraphics[width=0.47\textwidth]{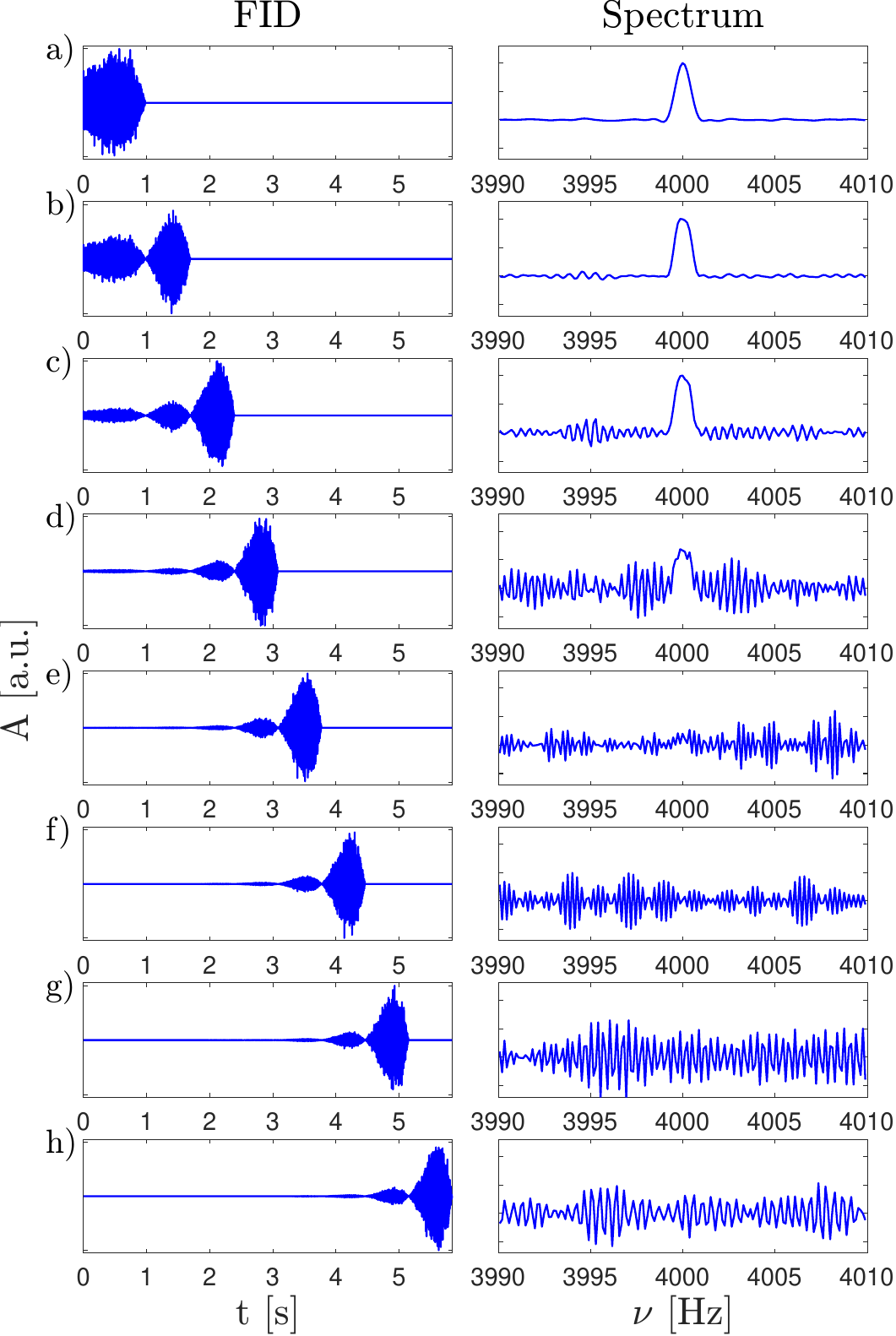}
\caption{\label{fig:RootNoise}Simulated FIDs and NMR-spectra using a single offset frequency $\nu = 4000$~Hz. The apodized time domain signal was truncated after (a) the first to (h) the eighth root of the applied window function (Eq.~\ref{eq:TDaFunc}, $\text{FWHM} = 1\:\text{Hz}$) and zero-filled, respectively, to achieve identical spectral resolutions. The desired parabolic line shapes can be obtained by truncation at the second root but the signal-to-noise ratio significantly decreases for later truncations. Amplitudes in FIDs and spectra are normalized with respect to the maximum intensity. Note that the FID signal gets more intense with every additional step which is due to stronger noise amplification. This leads to a point (e) where the spectral line fully vanishes and only noise remains.}
\end{figure}

\begin{figure}[H]
\centering
\includegraphics[width=0.41\textwidth]{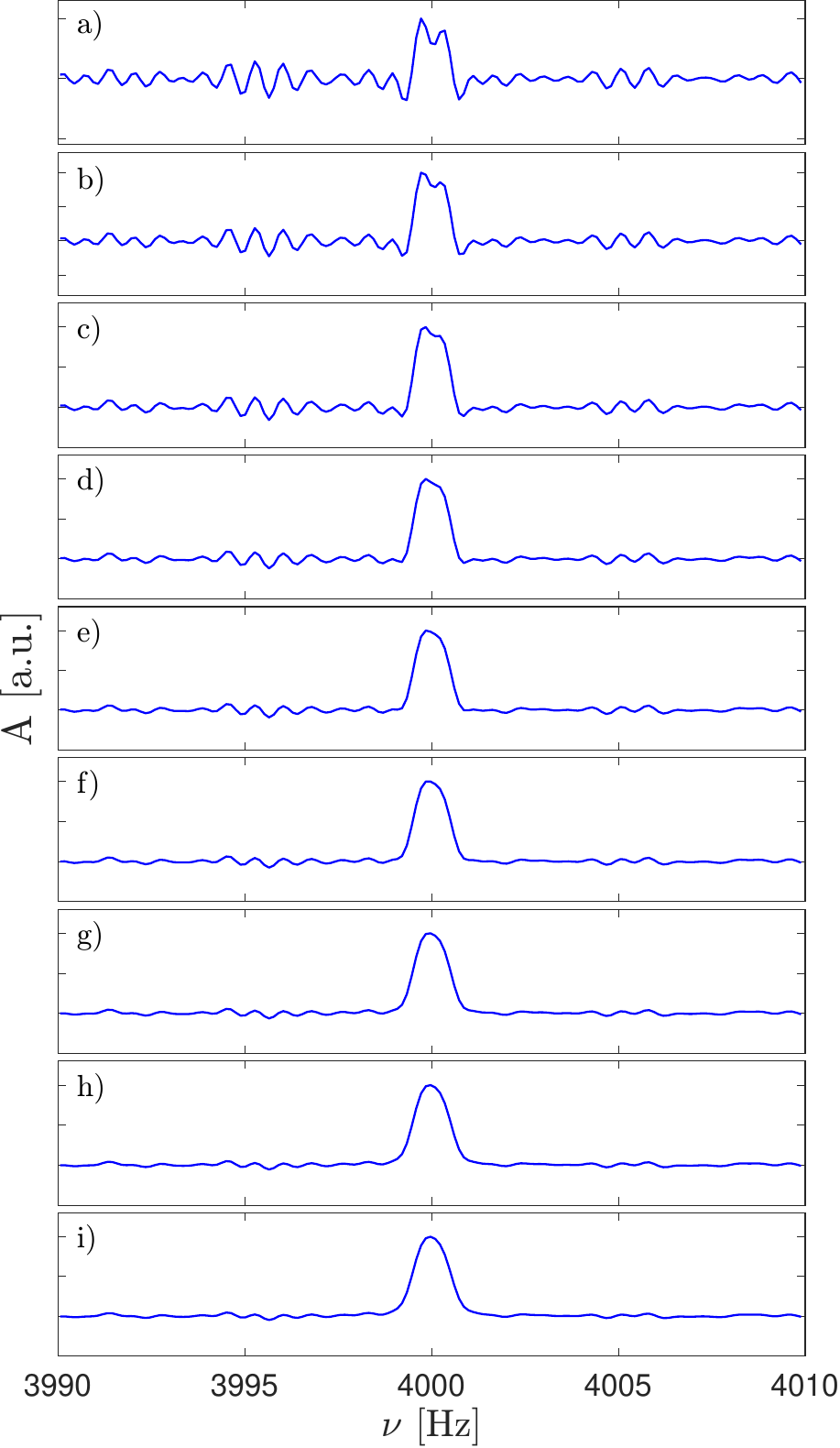}
\caption{\label{fig:CompNoise}Simulated spectra using a single offset frequency $\nu = 4000$~Hz. The original time signal decay corresponding to a damping constant $k = \frac{1}{T_2} = \big(\frac{1}{\pi} \text{s}\big)^{-1} \approx 3.14$~Hz was compensated by multiplying the signal with an exponential function $f_{\text{comp}} = \exp(k_{\text{comp}}t)$ using compensation constants $k_{\text{comp}} = T_{2,\text{comp}}^{-1}$ with $T_{2,\text{comp}}$-values ranging from $\frac{2}{3}T_2$ (a) to $\frac{4}{3}T_2$ (i) in steps of $\frac{1}{12}T_2$. Case e) thus corresponds to a perfectly compensated decay. All FIDs were apodized to give parabolic lines in the frequency domain with line width $\text{FWHM} = 1\:\text{Hz}$. The resulting lines possess Lorentzian contributions (undercompensation, cases f) to i)) or become distorted in overcompensation cases ( examples a) to d)). Due to noise amplification, FIDs were truncated at the second root of the window function and zero-filled.}
\end{figure}

\begin{figure}[H]
\centering
\includegraphics[width=0.485\textwidth]{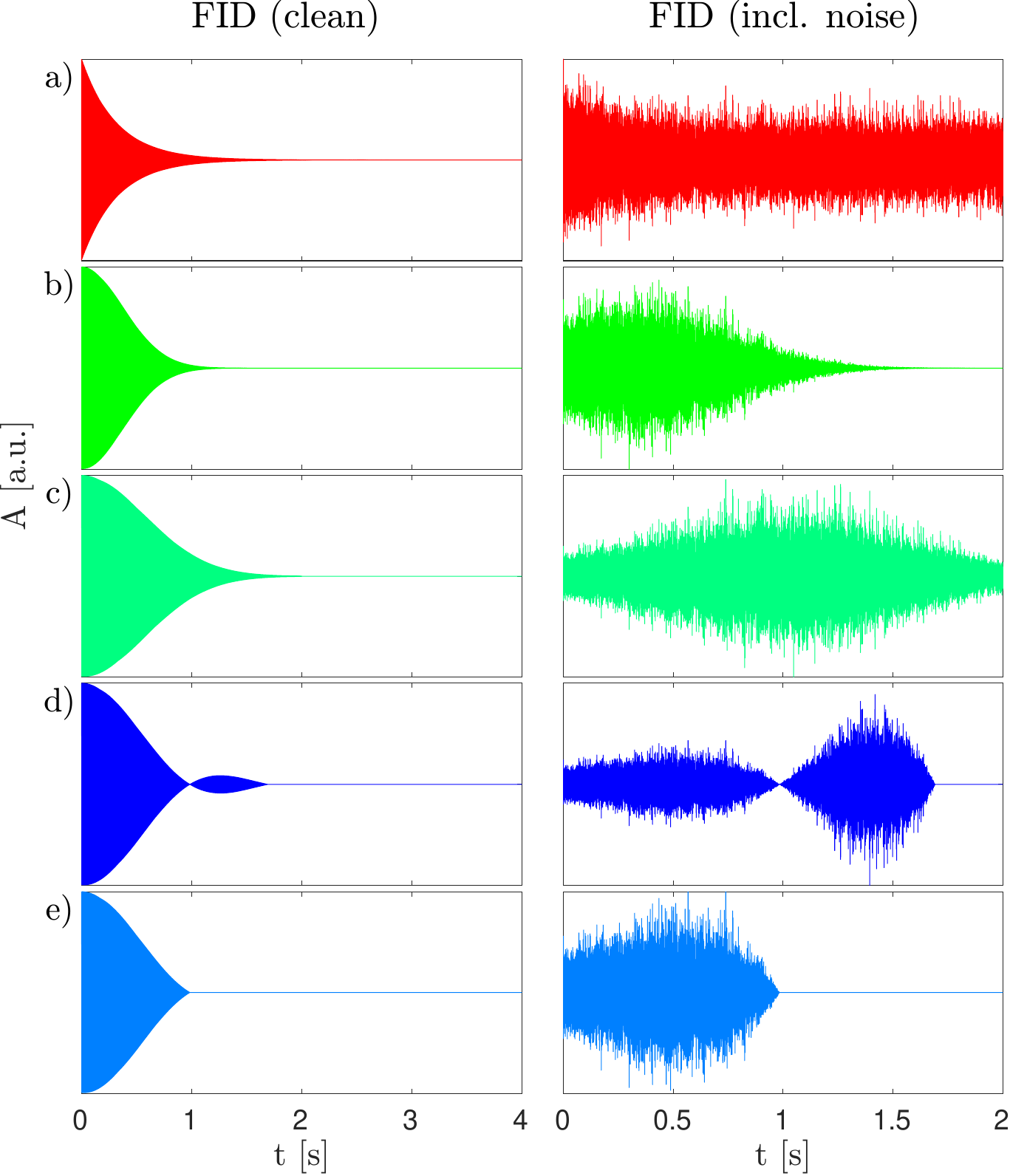}
\caption{\label{fig:FIDNoise}Simulated clean and noisy FIDs using a single offset frequency $\nu = 4000$~Hz. The original unchanged exponentially decaying time signal (a) has a damping constant of $\frac{1}{T_2} = \big(\frac{1}{\pi} \text{s}\big)^{-1}$.  The applied noise profile was adapted to give a signal-to-noise ratio (S/N) of 10 in the frequency domain and applied to each of the FIDs. Cases b) and c) represent a Lorentz-to-Gaussian apodization using different widths of the applied Gaussian window functions. The parabolic line shape apodization is shown in cases d) and e) where the signal was truncated at the second or first root of the window function to avoid heavy noise amplification, respectively. All FIDs were zero-filled before applying any Fourier transform (not shown). All amplitudes are normalized. Noisy FIDs were truncated after two seconds (cases a-e).}
\end{figure}

\begin{figure}[H]
\includegraphics[width=0.47\textwidth]{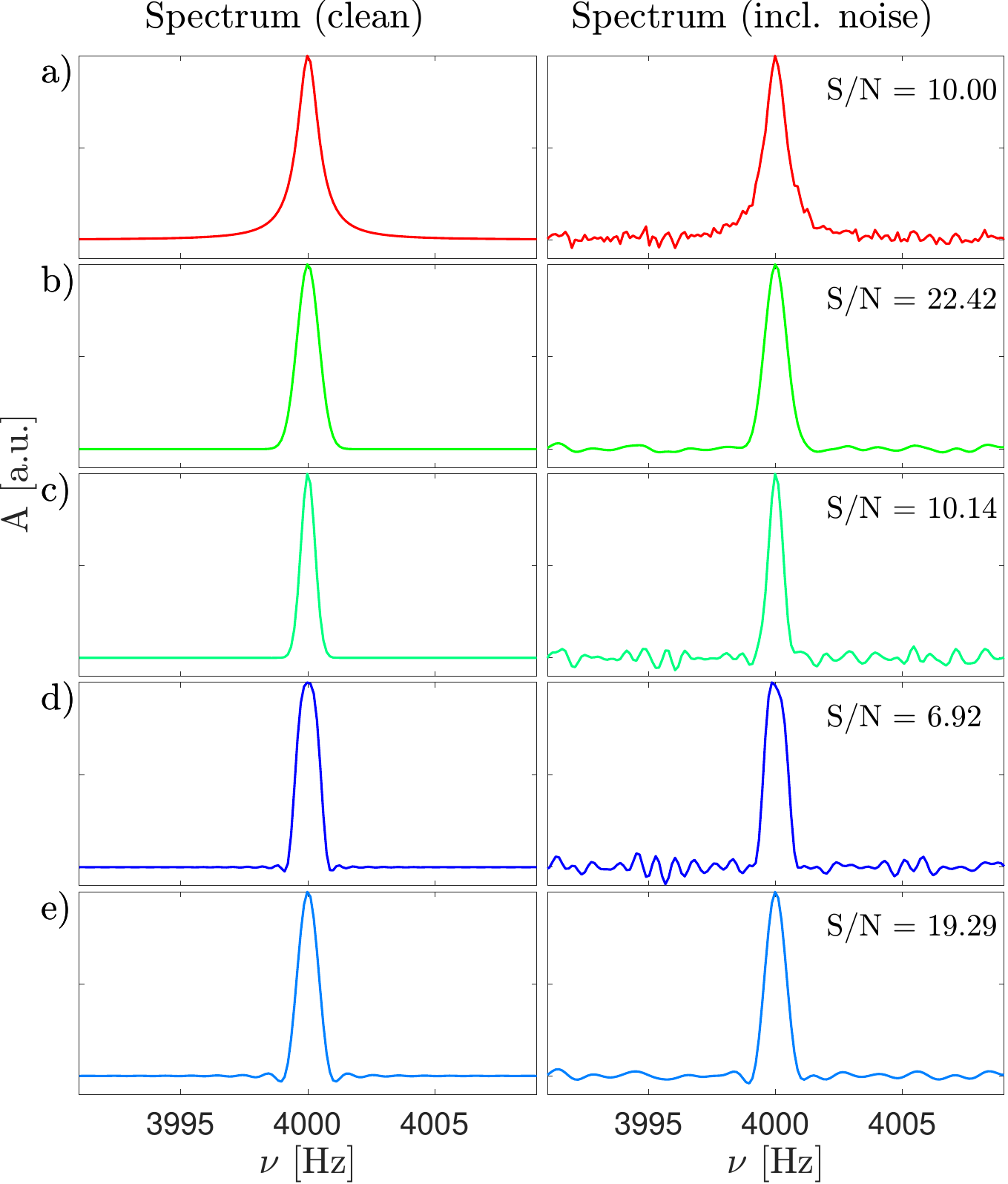}
\caption{\label{fig:SpecNoise}Simulated clean and noisy spectra obtained by Fourier transforming the signals depicted in FIG.~\ref{fig:FIDNoise}. The exponentially decaying time signal (a) gives a Lorentzian line with a signal-to-noise ratio (S/N) of 10. Cases b) and c) yield Gaussian lines. The narrower Gaussian (c) was adjusted to intersect the parabolic line at 3\% of the maximum amplitude, allowing for a better comparison due to inherently different support properties. The parabolic line apodization d) leads to a smaller signal-to-noise ratio. Truncation at the first window function root during the parabolic line apodization e) gives a higher signal-to-noise ratio but the line shape is significantly less parabolic.}
\end{figure}

First, it is possible to determine whether the window function amplifies noise. This effect frequently occurs due to the window function insufficiently counteracting the noise amplification that is introduced by the signal decay compensation prior to the window function application. As can be seen in FIG.~\ref{fig:RootNoise}, the noise is indeed amplified and thus requires truncation of the time signal at reasonable signal durations. In the given example, the signal needs to be truncated latest at the second root of the window function which, as was suggested in section~\ref{sec:ch5}, still allows to obtain a satisfying line shape whilst keeping noise at moderate levels.\newline
Analogously to the simulations presented in FIG.~\ref{fig:Comp} (see section~\ref{sec:ch5}), in presence of noise, lines possess Lorentzian contributions when Fourier transforming undercompensated signals whereas overcompensation distorts the spectral line and introduces additional noise. A corresponding simulated spectrum series including noise is shown in FIG.~\ref{fig:CompNoise}. As discussed previously, the time signal was truncated at the second root of the window function to avoid heavy noise amplification artifacts.
The obtained insights can further be used in a direct comparison of different line shapes achieved by dedicated apodization techniques. For this purpose, we  processed the same FID using the Lorentz-to-Gaussian and our parabolic line apodizations and compared the Fourier transformed spectra to that obtained for the non-apodized FID. The resulting FIDs are visualized in FIG.~\ref{fig:FIDNoise} and the corresponding spectra can be seen in FIG.~\ref{fig:SpecNoise}. Both, FIDs and spectra, are displayed clean (noiseless) and including noise.
Gaussian window functions and the parabolic line window function, when used for spectrum apodization, can lead to a reduction of the signal-to-noise ratio. Nonetheless, the possibility to truncate parabolic-line-apodized FIDs after only a few of the window function roots turns out to be useful for limiting noise amplification while achieving similar results as for the remaining apodization approaches.

Fourier transforming the obtained FIDs gives spectra with different signal-to-noise ratios. Indeed, both Gaussian apodizations result in improved or similar signal-to-noise ratios compared to the unapodized Lorentzian spectrum which in case (b) of Fig.~\ref{fig:SpecNoise} is due to noise dampening for large portions of the FID. Lorentz-to-Gaussian transforms are commonly used to achieve narrower lines such as shown in case (c) of the provided spectrum series. Note that such Gaussian apodizations generally decrease the signal-to-noise ratio depending on the present amount of noise in the original FID and the chosen linewidth.

The signal-to-noise ratio in the parabolic spectrum (d) is slightly worse than in the Gaussian and Lorentzian cases. Nonetheless, the obtained spectral line has a truncated parabolic shape. On the other hand, truncating at the first root of the parabolic line shape window function (e) improves the signal-to-noise ratio but leads to a significantly less parabolic shape that resembles the Gaussian profile in case (c). Our apodization approach is thus applicable even for noise-containing signals but requires careful handling to ensure appropriate outcomes.

\newpage
\clearpage
\bibliography{NewtonTransform}

\end{document}